\documentclass[12pt]{article}
\pdfoutput=1

\usepackage{amsmath,amsthm,amssymb,euscript,epsf,epsfig}
\usepackage{array,latexsym,amsfonts,bm,enumerate,url}

\usepackage{graphicx}
\usepackage[usenames]{color}
\usepackage[english]{babel}

\usepackage{fancybox}

\usepackage{mciteplus}

\usepackage{parskip}     % For a blank line between paragraphs & no indenting
\usepackage{titlesec}
\titlespacing{\section}{0pt}{3.5ex}{1ex}
\titlespacing{\subsection}{0pt}{2.5ex}{1ex}
\titlespacing{\subsubsection}{0pt}{2ex}{0ex}

\newcommand{\captionfonts}{\small}
\makeatletter  % Allow the use of @ in command names
\long\def\@makecaption#1#2{%
  \vskip\abovecaptionskip
  \sbox\@tempboxa{{\captionfonts #1: #2}}%
  \ifdim \wd\@tempboxa >\hsize
    {\captionfonts #1: #2\par}
  \else
    \hbox to\hsize{\hfil\box\@tempboxa\hfil}%
  \fi
  \vskip\belowcaptionskip}
\makeatother   % Cancel the effect of \makeatletter

\usepackage[
      colorlinks=true,
      linkcolor=blue,
      urlcolor=blue,
      filecolor=blue,
      citecolor=blue,
      pdfstartview=FitH,
      pdftitle={},
      pdfauthor={},
      pdfsubject={},
      pdfkeywords={},
      pdfpagemode=UseNone,
      bookmarksopen=true
      ]{hyperref}
\usepackage[all]{hypcap}     %should be loaded AFTER hyperref, which should otherwise be loaded last.

\begin{document}

\numberwithin{equation}{section}

%%%%%%%%%%%%%%%%%%%%%%%%%%%%%%%%%%%%%%%%%%%%%%%%%%%%%%%%%%%%
%                       DEFINITIONS

\mathchardef\mhyphen="2D

%%%%%%%%%%%%%%%%%%%%%%%%%%%%%%%%%%%%%%%%%%%%%%%%%%%%%%%%%%%%
%                        Commands

\newcommand{\be}{\begin{equation}}
\newcommand{\ee}{\end{equation}}
\newcommand{\bea}{\begin{eqnarray}\displaystyle}
\newcommand{\eea}{\end{eqnarray}}
\newcommand{\nnm}{\nonumber}
\newcommand{\nn}{\nonumber}

\def\eq#1{(\ref{#1})}
\newcommand{\secn}[1]{Section~\ref{#1}}

\newcommand{\tbl}[1]{Table~\ref{#1}}
\newcommand{\fig}{Fig.~\ref}

\def\beq{\begin{equation}}
\def\eeq{\end{equation}}
\def\beqa{\begin{eqnarray}}
\def\eeqa{\end{eqnarray}}
\def\bet{\begin{tabular}}
\def\eet{\end{tabular}}
\def\bs{\begin{split}}
\def\es{\end{split}}

%%%%%%%%%%%%%%%%%%%%%%%%%%%%%%%%%%%%%%%%%%%%%%%%%%%%%%%%%%%%%
%                        Greek letters

\def\a{\alpha}  
\def\b{\beta}  
\def\c{\chi}
\def\g{\gamma}
\def\G{\Gamma}
\def\e{\epsilon}
\def\vep{\varepsilon}
\def\tvep{\tilde{\varepsilon}}
\def\f{\phi}
\def\F{\Phi}
\def\fb{{\ov \phi}}
\def\vf{\varphi}
\def\m{\mu}
\def\mub{\ov \mu}
\def\n{\nu}
\def\nub{\ov \nu}
\def\o{\omega}
\def\O{\Omega}
\def\r{\rho}
\def\k{\kappa}
\def\kab{\ov \kappa}
\def\s{\sigma}
\def\t{\tau}
\def\th{\theta}
\def\sb{\ov\sigma}
\def\S{\Sigma}
\def\l{\lambda}
\def\L{\Lambda}
\def\p{\psi}

%%%%%%%%%%%%%%%%%%%%%%%%%%%%%%%%%%%%%%%%%%%%%%%%%%%%%%%%%%%%%
%              Calligraphic & Blackboard letters etc

\def\cA{{\cal A}} \def\cB{{\cal B}} \def\cC{{\cal C}}
\def\cD{{\cal D}} \def\cE{{\cal E}} \def\cF{{\cal F}}
\def\cG{{\cal G}} \def\cH{{\cal H}} \def\cI{{\cal I}}
\def\cJ{{\cal J}} \def\cK{{\cal K}} \def\cL{{\cal L}}
\def\cM{{\cal M}} \def\cN{{\cal N}} \def\cO{{\cal O}}
\def\cP{{\cal P}} \def\cQ{{\cal Q}} \def\cR{{\cal R}}
\def\cS{{\cal S}} \def\cT{{\cal T}} \def\cU{{\cal U}}
\def\cV{{\cal V}} \def\cW{{\cal W}} \def\cX{{\cal X}}
\def\cY{{\cal Y}} \def\cZ{{\cal Z}}

\def\mC{\mathbb{C}} 
\def\mP{\mathbb{P}}  
\def\mR{\mathbb{R}} 
\def\mZ{\mathbb{Z}} 
\def\mT{\mathbb{T}} 
\def\mN{\mathbb{N}}
\def\mH{\mathbb{H}}
\def\mX{\mathbb{X}}

\def\C{\mathbb{C}}
\def\CP{\mathbb{CP}}
\def\R{\mathbb{R}}
\def\RP{\mathbb{RP}}
\def\Z{\mathbb{Z}}
\def\N{\mathbb{N}}
\def\H{\mathbb{H}}

\newcommand{\rmd}{\mathrm{d}}
\newcommand{\rmx}{\mathrm{x}}

\def\tA{ {\widetilde A} } 

\def\one{{\hbox{\kern+.5mm 1\kern-.8mm l}}}
\def\zero{{\hbox{0\kern-1.5mm 0}}}

%%%%%%%%%%%%%%%%%%%%%%%%%%%%%%%%%%%%%%%%%%%%%%%%%%%%%%%%%%%%%%
%                                Colours

\newcommand{\red}[1]{{\color{red} #1}}
\newcommand{\blue}[1]{{\color{blue} #1}}
\newcommand{\green}[1]{{\color{green} #1}}

\definecolor{orange}{rgb}{1,0.5,0}
\newcommand{\orange}[1]{{\color{orange} #1}}

%%%%%%%%%%%%%%%%%%%%%%%%%%%%%%%%%%%%%%%%%%%%%%%%%%%%%%%%%%%%%%
%                                 QM               

\newcommand{\bra}[1]{{\langle {#1} |\,}}
\newcommand{\ket}[1]{{\,| {#1} \rangle}}
\newcommand{\braket}[2]{\ensuremath{\langle #1 | #2 \rangle}}
\newcommand{\Braket}[2]{\ensuremath{\langle\, #1 \,|\, #2 \,\rangle}}
\newcommand{\norm}[1]{\ensuremath{\left\| #1 \right\|}}
\def\corr#1{\left\langle \, #1 \, \right\rangle}
\def\vac{|0\rangle}

\newcommand{\com}[2]{[#1,\, #2]}

%%%%%%%%%%%%%%%%%%%%%%%%%%%%%%%%%%%%%%%%%%%%%%%%%%%%%%%%%%%%%
%                         General

\def\d{ \partial } 
\def\zb{{\bar z}}

\newcommand{\sq}{\square}
\newcommand{\IP}[2]{\ensuremath{\langle #1 , #2 \rangle}}    %inner product

\newcommand{\floor}[1]{\left\lfloor #1 \right\rfloor}
\newcommand{\ceil}[1]{\left\lceil #1 \right\rceil}

\newcommand{\dbyd}[1]{\ensuremath{ \frac{\d}{\d {#1}}}}
\newcommand{\ddbyd}[1]{\ensuremath{ \frac{\d^2}{\d {#1}^2}}}

\newcommand{\Zd}{\ensuremath{ Z^{\dagger}}}
\newcommand{\Xd}{\ensuremath{ X^{\dagger}}}
\newcommand{\Ad}{\ensuremath{ A^{\dagger}}}
\newcommand{\Bd}{\ensuremath{ B^{\dagger}}}
\newcommand{\Ud}{\ensuremath{ U^{\dagger}}}
\newcommand{\Td}{\ensuremath{ T^{\dagger}}}

\newcommand{\T}[3]{\ensuremath{ #1{}^{#2}_{\phantom{#2} \! #3}}}		%general tensor with upper indices displayed first 

\newcommand{\tr}{\operatorname{tr}}
\newcommand{\Str}{\operatorname{Str}}
\newcommand{\sech}{\operatorname{sech}}
\newcommand{\Spin}{\operatorname{Spin}}
\def\Tr{{\rm Tr\, }} 
\newcommand{\Sym}{\operatorname{Sym}}
\newcommand{\Com}{\operatorname{Com}}
\def\adj{\textrm{adj}}
\def\id{\textrm{id}}
\def\Id{\textrm{Id}}
\def\ind{\textrm{ind}}
\def\Dim{\textrm{Dim}}
\def\End{\textrm{End}}
\def\Res{\textrm{Res}}
\def\Ind{\textrm{Ind}}
\def\ker{\textrm{ker}}
\def\im{\textrm{im}}
\def\sgn{\textrm{sgn}}
\def\ch{\textrm{ch}}
\def\STr{\textrm{STr}}
\def\Sym{\textrm{Sym}}

\def\ha{\frac{1}{2}}
\def\tha{\tfrac{1}{2}}
\def\wt{\widetilde}
\def\ra{\rangle}
\def\la{\langle}

\def\pb{\ov\psi}
\def\pt{\widetilde{\psi}}
\def\at{\widetilde{\a}}
\def\cb{\ov\chi}
\def\d{\partial}
\def\db{\bar\partial}
\def\delb{\bar\partial}
\def\dbar{\ov\partial}
\def\dag{\dagger}
\def\dalpha{{\dot\alpha}}
\def\dbeta{{\dot\beta}}
\def\dgamma{{\dot\gamma}}
\def\ddelta{{\dot\delta}}
\def\ad{{\dot\alpha}}
\def\bd{{\dot\beta}}
\def\dg{{\dot\gamma}}
\def\dd{{\dot\delta}}
\def\th{\theta}
\def\Th{\Theta}
\def\eb{{\ov \epsilon}}
\def\gb{{\ov \gamma}}
\def\wb{{\ov w}}
\def\Wb{{\ov W}}
\def\ib{{\ov i}}
\def\jb{{\ov j}}
\def\kb{{\ov k}}
\def\mb{{\ov m}}
\def\nb{{\ov n}}
\def\qb{{\ov q}}
\def\Qb{{\ov Q}}
\def\xh{\hat{x}}
\def\D{\Delta}
\def\DD{\Delta^\dag}
\def\Db{\ov D}
\def\M{{\cal M}}
\def\rd{\sqrt{2}}
\def\ov{\overline}
\def\Slash{\, / \! \! \! \!}
\def\dslash{\partial\!\!\!/} 
\def\Dslash{D\!\!\!\!/\,\,}
\def\fslash#1{\slash\!\!\!#1}
\def\Fslash#1{\slash\!\!\!\!#1}

\def\del{\partial}
\def\delb{\bar\partial}
\newcommand{\ex}[1]{{\rm e}^{#1}} 
\def\ii{{i}}

\renewcommand{\theequation}{\thesection.\arabic{equation}}
\newcommand{\vs}[1]{\vspace{#1 mm}}

\newcommand{\ve}{{\vec{\e}}}
\newcommand{\shalf}{\frac{1}{2}}

\newcommand{\lb}{\rangle}
\newcommand{\al}{\ensuremath{\alpha'}}
\newcommand{\ap}{\ensuremath{\alpha'}}

\newcommand{\bean}{\begin{eqnarray*}}
\newcommand{\eean}{\end{eqnarray*}}
\newcommand{\ft}[2]{{\textstyle {\frac{#1}{#2}} }}

\newcommand{\hsp}{\hspace{0.5cm}}
\def\half{{\textstyle{1\over2}}}
\let\ci=\cite \let\re=\ref
\let\se=\section \let\sse=\subsection \let\ssse=\subsubsection

\newcommand{\dpb}{D$p$-brane}
\newcommand{\dpbs}{D$p$-branes}

\def\gh{{\rm gh}}
\def\sgh{{\rm sgh}}
\def\NS{{\rm NS}}
\def\R{{\rm R}}
\def\Qp{Q_{\rm P}}
\def\QP{Q_{\rm P}}

\newcommand\dott[2]{#1 \! \cdot \! #2}

\def\eo{\overline{e}}

%\renewcommand{\arraystretch}{2}

%%%%%%%%%%%%%%%%%%%%%%%%%%%%%%%%%%%%%%%%%%%%%%%%%%%%%%%%%%%%%%%%%%%%%%%%%%%%%%

%\renewcommand{\theequation}{\arabic{section}.\arabic{equation}}
\def\p{\partial}
\def\h{{1\over 2}}

\def\d{\partial}
\def\la{\lambda}
\def\eps{\epsilon}
\def\bb{\bigskip}
\def\mm{\medskip}
\def\tg{\tilde\gamma}
\newcommand{\dm}{\begin{displaymath}}
\newcommand{\edm}{\end{displaymath}}
\renewcommand{\b}{\tilde{B}}
\newcommand{\gm}{\Gamma}
\newcommand{\ac}[2]{\ensuremath{\{ #1, #2 \}}}
\renewcommand{\ell}{l}
\newcommand{\z}{\ell}
\newcommand{\newsection}[1]{\section{#1} \setcounter{equation}{0}}
\def\bb{$\bullet$}
\def\Qbar{{\bar Q}_1}
\def\QPbar{{\bar Q}_p}

\def\q{\quad}

\def\bn{B_\circ}

\let\a=\alpha \let\b=\beta \let\g=\gamma \let\d=\delta \let\e=\epsilon
\let\c=\chi \let\th=\theta  \let\k=\kappa
\let\l=\lambda \let\m=\mu \let\n=\nu \let\x=\xi \let\r=\rho
\let\s=\sigma \let\t=\tau
\let\vp=\varphi \let\vep=\varepsilon
\let\w=\omega      \let\G=\Gamma \let\D=\Delta \let\Th=\Theta
                     \let\P=\Pi \let\S=\Sigma

\def\h{{1\over 2}}
\def\t{\tilde}
\def\r{\rightarrow}
\def\nn{\nonumber\\}
\let\bm=\bibitem
\def\Kt{{\tilde K}}
\def\b{\bigskip}

\let\p=\partial

%%%%%%%%%%%%%%%%%%%%%%%%%%%%%%%%%%%%%%%%%%%%%%%%%%%%%%%%%%%%%%%%%%%%%%%%5

\begin{flushright}
%OHSTPY-HEP-T-03-012\\
\end{flushright}
\vspace{20mm}

\begin{center}
{\LARGE  Microstates at the boundary of AdS}
\\
\vspace{2cm}

{\bf Samir D. Mathur ~and~ David Turton} \\

\vspace{15mm}
Department of Physics,\\ The Ohio State University,\\ Columbus,
OH 43210, USA

\vspace {5mm}
mathur@mps.ohio-state.edu\\
turton.7@osu.edu
\vspace{4mm}

\end{center}

\vspace{1cm}

\begin{abstract}

The bound states of the D1D5 brane system have a known gravitational description: flat asymptotics,  an anti-de Sitter region,  and a `cap'  ending the AdS region. We construct perturbations that correspond to the action of chiral algebra generators on Ramond ground states of  D1D5 branes. Abstract arguments in the literature suggest that the perturbation should be pure gauge in the AdS region; our perturbation indeed has this structure, with the nontrivial deformation of the geometry occurring at the `neck'  between the AdS region and asymptotic infinity. This `non-gauge' deformation is needed to provide the nonzero energy and momentum carried by the perturbation. We also suggest implications this structure may have for the majority of microstates which live at the cap.

\end{abstract}

\thispagestyle{empty}

\newpage

\section{Introduction}

Black holes have an entropy proportional to their surface area \cite{Bekenstein:1973ur,*Hawking:1974sw}, suggesting that the states corresponding to this entropy should be localized around the horizon~\cite{Bekenstein:1980jp}. 
But in ordinary gravity theories we typically find no structure at the horizon -- black holes `have no hair'~\cite{Israel:1967wq,*Carter:1971zc,*Price:1971fb,*Price:1972pw,*Robinson:1975bv,*Heusler:1998ua}, a fact which leads to the information paradox~\cite{Hawking:1976ra}, see also \cite{Mathur:2009hf,*Mathur:2011uj}.

For black holes with a near-horizon $AdS$ region, gauge/gravity duality~\cite{Maldacena:1997re,Gubser:1998bc,Witten:1998qj} allows one to replace the entire $AdS$ region by a CFT placed at the boundary of the $AdS$. 
In $AdS_5$ one encounters a separation between bulk degrees of freedom and a $U(1)$ degree of freedom localized at the boundary of AdS~\cite{Witten:1998qj}. 
So one may ask: in the gravity description, which states (if any) are localized at the boundary of $AdS$, at the horizon, or deep inside the black hole?

We address this question in $AdS_3$, where it has been argued that the boundary diffeomorphisms of $AdS_3$ form a Virasoro algebra with a central charge $c$~\cite{Brown:1986nw}. It has been further argued that this $c$ should give the number of states in the 2D boundary theory at energy level $N$ by the standard expression $\exp[2\pi\sqrt{cN\over 6}]$ \cite{Strominger:1997eq}.
Carlip has extended these ideas to general black hole horizons, and proposed ideas relating conformal symmetry to thermodynamic properties of black holes~\cite{Carlip:1998wz,*Carlip:1999cy,*Carlip:1999db}.

We study the D1D5 bound state of string theory, which generates a near-horizon $AdS_3$ geometry and yields a black hole in $AdS_3$~\cite{Strominger:1996sh,*Callan:1996dv,*Banados:1992wn} when excitation energy is added. Suppose we start with an extremal state $|\psi\rangle$ of the D1D5 CFT which admits a classical gravitational description. Such geometries have the following schematic structure: 

\begin{figure}[htbp]
\begin{center}
\includegraphics[width=11cm]{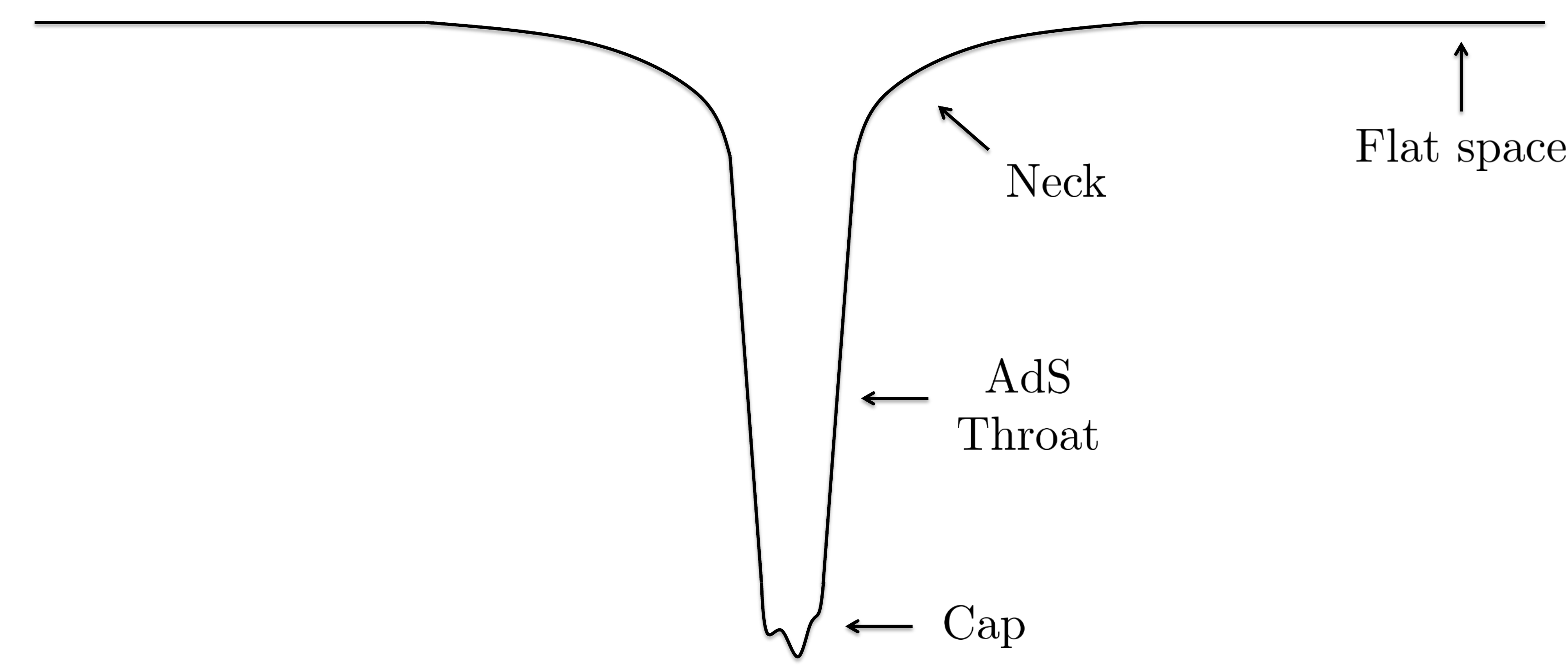}
\caption{Structure of D1D5 geometries and associated terminology}
\label{default}
\end{center}
\end{figure}

There is flat space at infinity, then a `neck' region which leads to a $AdS_3$ `throat'. The throat ends in a `cap' 
which depends on the state $|\psi\rangle$
\cite{Lunin:2001fv,Lunin:2002iz,*Kanitscheider:2007wq,*Skenderis:2008qn,
Bena:2004de,*Bena:2005va,*Berglund:2005vb,*Saxena:2005uk,*Bena:2007kg,*Das:2008ka}. 
This structure has also been seen for non-extremal states
\cite{Jejjala:2005yu,*Chowdhury:2007jx,*Avery:2009tu,*Mathur:2008nj}; for recent related work see
\cite{
Balasubramanian:2005mg,*Balasubramanian:2007bs,*Simon:2011zza,
Bena:2010gg,*Bena:2011uw,*Bena:2011zw,
Giusto:2011fy,*Giusto:2009qq,*Black:2010uq}.

\newpage

In this paper we construct the perturbation corresponding to the state 
\be
L_{-n}|\psi\rangle \nnm
\ee
where $L_{-n}$ is a Virasoro generator. The perturbed solution has D1, D5 and P charges. We find that to leading order the perturbation is a pure diffeomorphism in the $AdS$ throat,
but there is a genuine deformation at the neck region, which forms a natural boundary to the $AdS$ throat. 
The perturbation: 
\begin{enumerate}[(i)]
	\item reduces to being a pure diffeomorphism in the $AdS$ throat+cap;
	\item is {\it not} pure gauge in the neck;
	\item is normalizable at infinity;
	\item is regular in the `cap' region;
	\item has the correct spacetime dependence to carry the energy and momentum \\ required by the excitation $L_{-n}$.
\end{enumerate}
Thus our construction is consistent with the notion that in $AdS_3$ the action of the Virasoro algebra is represented by `boundary diffeomorphisms'~\cite{Brown:1986nw}. 
Note that the perturbation could not have been a pure diffeomorphism everywhere, since the state $L_{-n}|\psi\rangle$ has more energy than the state $|\psi\rangle$. \\

\begin{figure}[htbp]
\begin{center}
\includegraphics[width=15cm]{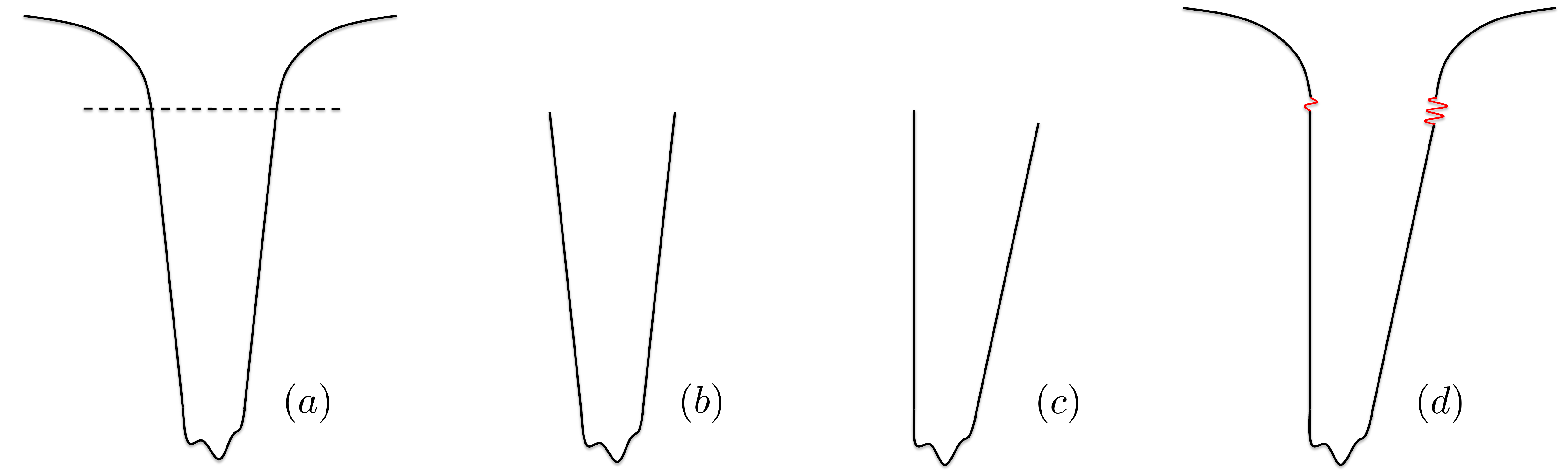}
\caption{Cutting the $AdS$ part of the geometry, performing a diffeomorphism and gluing it back, creating a real (i.e. not pure gauge) perturbation in the neck.}
\label{fig2}
\end{center}
\end{figure}

A cartoon of how one may imagine the perturbation is given in Fig.~\ref{fig2}. Part $(a)$ depicts the entire geometry, with a dashed line indicating the location where the boundary CFT will be placed. In $(b)$ we truncate to the asymptotically $AdS$ region that is dual to the CFT. In $(c)$ we perform a diffeomorphism of region $(b)$, corresponding to the application of a Virasoro generator. In $(d)$ we glue this region back to flat space, creating a deformation in the `neck' region.

\newpage

In this paper we work in type IIB string theory compactified on $S^1\times T^4$. We wrap $n_1$ D1 branes on $S^1$ and $n_5$ D5 branes on $S^1\times T^4$, and so the full throat geometry is $AdS_3\times S^3\times T^4$.
In addition to constructing the $L_{-n}$ perturbation, we also carry out similar constructions for the left-moving $SU(2)$ current algebra generator $J^{(3)}_{-n}$ of the $S^3$ and the $U(1)$ currents $J^i_{-n}$ which arise from the translations along the $T^4$.

We find our perturbation by solving the equations of motion separately in the outer region (throat+neck+flat infinity) and inner region (throat+cap), and then matching them to leading order in the throat. Such a matching procedure was carried out to several orders for a different state in~\cite{Mathur:2003hj}. 

For the case of the $U(1)$ currents $J^i_{-n}$, the solutions in the outer region can be seen to be a limit of the solutions in~\cite{Larsen:1995ss,*Cvetic:1995bj,*Horowitz:1996th,*Horowitz:1996cj}, where waves were added to a black string geometry. These latter solutions turned out to be singular at the horizon of the black string~\cite{Kaloper:1996hr,*Horowitz:1997si}. Such a singularity was encountered again in the construction of `external hair' on extremal holes~\cite{Banerjee:2009uk,Jatkar:2009yd}, and it appears reasonable to attribute such singularities to the general `no-hair theorem' which forbids smooth perturbations to a horizon. In the present work our geometry has a regular cap in place of a horizon, and so our perturbations do not become singular anywhere. We shall comment further on this issue in the discussion section.

The number of states generated by the chiral algebra is small, corresponding to a $U(1)$ degree of freedom with central charge $c=6$. 
The remaining degrees of freedom, corresponding to a central charge $c=6(n_1n_5-1)$, are expected to live at the cap. 
Our construction of states at the neck suggests a construction of states localized in the cap, which we discuss in Section \ref{sec:Discussion}.

This paper is organized as follows:

\begin{itemize}
	\item In Section \ref{sectwo} we review the 2-charge background solution, the equations to be solved, and describe the perturbative matching process.
	\item In Section \ref{sec:Ln} we derive the perturbations for the $L_{-n}$ generators by solving the equations of motion in the neck and cap, and show that they agree in their domain of overlap.
	\item Sections \ref{sec:Jn} and \ref{sec:Jz} do the same for perturbations of the $S^3$ and $T^4$ respectively.
	\item Section \ref{sec:Check} gives a cross-check of our results against a test of $AdS$/CFT.
	\item Section \ref{sec:Discussion} discusses our results.
	\item Some technical details are placed in the appendices.
\end{itemize}

\newpage

\section{The 2-charge solution and the perturbation} \label{sectwo}

\subsection{The 2-charge solution} \label{sec:background_fields}

We take type IIB string theory with the compactification
\be
M_{9, 1}~\r~ M_{4, 1}\times S^1\times T^4 \,.
\ee
We wrap $n_1$ D1 branes on $S^1$ and $n_5$ D5 branes on $S^1\times T^4$. This D1D5 system can be mapped by dualities to the NS1P system, where the D5 branes become a multiwound fundamental string along the $S^1$ and the D1 branes become momentum P carried as travelling waves on the string. The gravitational solution of such a string carrying waves is known~\cite{Garfinkle:1990jq,*Callan:1995hn,*Dabholkar:1995nc,Lunin:2001fv}, and may be dualized back to obtain gravitational solutions describing the D1D5 bound states~\cite{Lunin:2001fv,Lunin:2001jy}. We use light-cone coordinates
\be
v = t-y \,, \qquad u = t+y 
\ee
constructed from the time $t$ and $S^1$ coordinate $y$. The D1D5 solutions are given in terms of a profile function $\vec F(v)$ describing the vibration profile of the NS1 in the NS1P duality frame. Here $\vec F$ is the transverse displacement of the vibrating string and the dependence on $v$  tells us that the waves move only in one direction along the string, generating a BPS state. 

The generic state in this system is quantum in nature and the gravitational field it sources is not well-described by a classical geometry. This is simply because the momentum on the NS1 is spread over many different harmonics, with an occupation number of order unity per harmonic. But we can gain insight into the physics of generic solutions by starting with special ones, where all the momentum is placed in a few harmonics. In this case we can take coherent states for the vibrating NS1 string, which are now described by a classical profile function $\vec F(v)$ and which generate well-described  classical geometries~\cite{Mathur:2005zp}. We use the simplest cap geometry, which is obtained for the choice of profile function of the form
\be \label{eq:helix_profile}
F_1(v)=a\cos{v\over n_5 R_y}, \quad F_2(v)=a\sin{v\over n_{5} R_y}, \quad
F_3(v)=0, \quad F_4(v)=0 \,,
\ee
where $R_y$ is the radius of the $S^1$.
The geometry for this choice of $\vec F$ had arisen earlier in different studies in \cite{Cvetic:1996xz,Balasubramanian:2000rt,Maldacena:2000dr}. 
The full solution is:
\bea
ds^2&=&-{1\over h}\left(dt^2-dy^2\right)+hf \left(d\theta^2+{dr^2\over r^2+a^2} \right) \cr
&& {}+h \left[ \left( r^2+{a^2Q^2\cos^2\theta\over h^2f^2} \right) \cos^2\theta d\psi^2
+\left(r^2+a^2-{a^2Q^2\sin^2\theta\over h^2f^2} \right)\sin^2\theta d\phi^2 \right] \cr
&& {} -{2aQ\over hf} \left(\cos^2 \theta dy d\psi+\sin^2\theta dt d\phi\right) +\sum_i R_i^2 dz_i^2 \,, \label{el} \\
C^{(2)}_{ty}&=&-{Q\over Q+f}  \,,
\qquad \qquad ~
C^{(2)}_{t\psi} ~=~ -{Qa\cos^2\theta\over Q+f}  \,,  \nonumber\\
C^{(2)}_{y\phi}&=&-{Qa\sin^2\theta\over Q+f} \,, 
\qquad\quad 
C^{(2)}_{\phi\psi} ~=~ Q\cos^2\theta+{Qa^2\sin^2\theta\cos^2\theta\over Q+f}   \label{selfdualfield}
\eea
where
\be
a={Q\over R_y}, \qquad f=r^2+a^2\cos^2\theta, \qquad h=1+{Q\over f} \,.
\ee

We take $R_y$ to be large compared to the curvature radius $\sqrt{Q}$ of the cap:
\be
\epsilon ~=~ {\sqrt{Q}\over R_y} ~\ll~ 1 \,.
\label{limitq}
\ee
More precisely, we assume $\e$ is small enough that we get several units of the $AdS$ radius between the cap and the neck. As an aside, to see how small $\e$ should be we note that the radial part of the metric is
\be
ds^2 ~\approx~ Q{dr^2\over r^2} ~=~ Q(d\log r)^2 \,.
\ee
The range of the $AdS$ throat region is from $r\sim a$ to $r\sim \sqrt{Q}$. Each proper length $\sqrt{Q}$ in the $r$ direction accounts for one $AdS$ radius. We then get $N_{AdS}$ units of the $AdS$ radius between the cap and the neck, where
\be
N_{AdS} ~\sim~ \log {\sqrt{Q}\over a} ~=~ \log {R_y\over \sqrt{Q}} ~=~ \log {1\over \epsilon} \gg 1
\ee
for sufficiently small $\e$. Thus if keep only the region $r \ll \sqrt{Q}$, we get a space that is asymptotically $AdS_3 \times S^3 \times T^4$. A similar result follows for all the D1D5 geometries in the limit (\ref{limitq}) \cite{Lunin:2002bj}. Thus we can associate a state in the dual CFT with each choice of cap, and also ask about the action of CFT operators $L_{-n}, J^a_{-n}, J^i_{-n}$ on the state. 

Let us consider the different parts of the geometry (\ref{el}); this structure is generic for two-charge geometries:

\begin{enumerate}[(i)]
	\item For $r\gg \sqrt{Q}$ we have approximately flat space.
	\item At $r\sim \sqrt{Q}$ we have the intermediate `neck' region.
	\item For $a \ll r \ll \sqrt{Q}$ we have the `throat', which is the Poincare patch of $AdS_3 \times S^3 \times T^4$.
	\item At $r\sim a$ we have the `cap', which depends on the choice of the state. 
\end{enumerate}

The geometry (\ref{el}) describes the particular state $|0\rangle_R$ in the Ramond sector of the D1D5 CFT which arises from the spectral flow~\cite{Schwimmer:1986mf} of the  NS vacuum $|0\rangle_{NS}$.

\subsection{The outer and inner regions of the background geometry}

Our goal is to construct perturbations on the geometry (\ref{el}) that correspond to states such as $L_{-n}|0\rangle_R$. We do this by splitting the geometry into two overlapping regions: an `outer region' and an `inner region'. We solve the equation for the perturbation in these two regions, and match the two solutions in the domain of overlap.

\begin{figure}[htbp]
\begin{center}
\includegraphics[width=14cm]{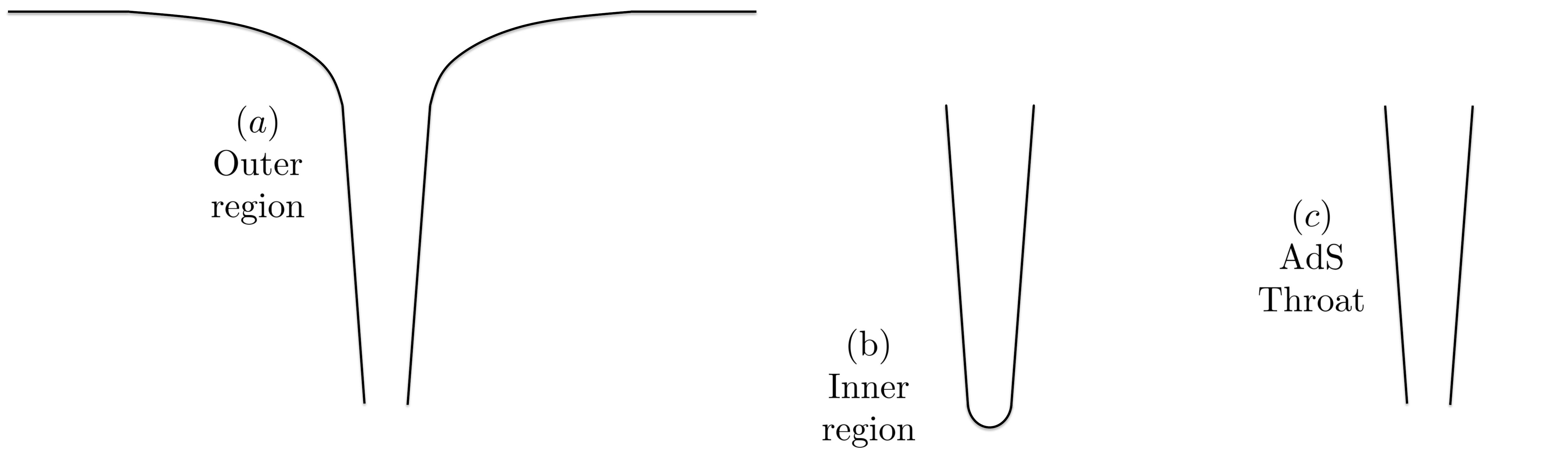}
\caption{The outer and inner regions, which overlap in the throat region.}
\label{fig:3}
\end{center}
\end{figure}

\subsubsection{The outer region}

The outer region is defined as
\be
r\gg a \,.
\ee
The solution (\ref{el}), \eq{selfdualfield} reduces in this limit to the `naive' D1D5 solution, depicted in Fig.\;\ref{fig:3}\,$(a)$:
\be
ds^2_{outer}= \left( 1+{Q\over r^2} \right)^{-1}\left[-dt^2+dy^2\right]
+\left(1+{Q\over r^2}\right)\left[dr^2+r^2d\Omega_3^2\right]+ \sum\limits_i R_i^2 dz_i^2 \,,
\label{outerr}
\ee
\be
C^{(2)}_{ty}=-{Q\over Q+r^2}, \quad C^{(2)}_{\psi\phi}=-Q\cos^2\theta \,.
\label{outerf}
\ee

\subsubsection{The inner region}

The inner region, depicted in Fig.\;\ref{fig:3}\,$(b)$, is defined by taking the limit
\be
r\ll \sqrt{Q} \,.
\ee
In this limit the geometry (\ref{el}) becomes
\bea
ds^2&=&-{(r^2+a^2\cos^2\theta)\over
Q}(dt^2-dy^2)+Q(d\theta^2+{dr^2\over r^2+a^2}) +Q(\cos^2\theta
d\psi^2+\sin^2\theta d\phi^2) \nonumber\\
&&\quad 
-2a(\cos^2\theta dyd\psi+\sin^2\theta dtd\phi)+\sum\limits_i R_i^2 dz_i^2 \,.
\label{innerr}
\eea
The change of coordinates
\be
\psi_{NS}=\psi-{a\over Q}y, \quad \phi_{NS}=\phi-{a\over Q}t
\label{spectral}
\ee
brings (\ref{innerr}) to the form $AdS_3\times S^3\times T^4$
\bea
ds^2 &=& -{(r^2+a^2)\over Q}dt^2+{r^2\over Q}dy^2+Q{dr^2\over r^2+a^2} \cr
&& {}+Q(d\theta^2+\cos^2\theta d\psi_{NS}^2+\sin^2\theta
d\phi_{NS}^2)+\sum\limits_i R_i^2 dz_i^2 \,.
\label{innerns}
\eea

\subsubsection{The overlap region (throat)}

The outer and inner regions overlap in the throat region, defined by
\be
a \ll r \ll \sqrt{Q} \,,
\ee
and depicted in Fig.\;\ref{fig:3}\,$(c)$. This domain of overlap exists because we have chosen the limit $\e \ll 1$ (\ref{limitq}).
The metric in this region can be obtained by taking the $r \ll \sqrt{Q}$ limit of the outer region solution \eq{outerr}, which gives
\bea
ds^2_{thr}&=&{r^2\over Q} \left[-dt^2+dy^2 \right]+{Q\over r^2} \left[dr^2+r^2d\Omega_3^2 \right] 
+ \sum\limits_i R_i^2 dz_i^2  \,, \nn
C^{(2)}_{ty}&=&{r^2\over Q} \,, \quad C^{(2)}_{\psi\phi}=-Q\cos^2\theta 
\label{adsg}
\eea
where we have added a constant unity to $C^{(2)}_{ty}$ (this does not affect the field strength). This geometry describes the Poincare patch of $AdS_3\times S^3\times T^4$.

\subsection{The field equations satisfied by the perturbation}

The 2-charge solution (\ref{el}), \eq{selfdualfield} is a 10D solution.  But since the $T^4$ has a constant size, dimensional reduction to 6D gives just the metric (\ref{el}) without the torus directions.  In this 6D solution, we find that the gauge field is self-dual
\be
*_6 F^{(3)}=F^{(3)}
\ee
(We use the sign convention $\epsilon_{tyr\theta\psi\phi}=\sqrt{-g_6}$.)

For the perturbations corresponding to $L_{-n}, J^a_{-n}$ we will take an ansatz that keeps the perturbed $F^{(3)}$ self-dual in the 6D space. The reason for this ansatz is that the 6D theory is described by a set of self-dual fields $F^r_S$ and anti-self-dual fields $F^i_A$, with the indices $r, i$ belonging to different internal symmetry groups. The background geometry has only one self-dual field turned on, and there is no natural pairing of any this field with any of the anti-self-dual fields. The operators $L_{-n}, J^a_{-n}$ do not carry any of the indices $r, i$, so we can expect that only the self-dual field making the background will be involved in the perturbations corresponding to these operators.

For self-dual $F^{(3)}$ we get $F^2\equiv F^{(3)}_{ABC}F^{(3)ABC}=0$. If the only contribution to the stress tensor is from $F^{(3)}$, then we get vanishing of the Ricci scalar $R$. The dilaton can be set to zero. The relevant fields are the metric $g_{AB}$ in 6D and the self dual gauge field $F^{(3)}_{ABC}$. We write $(A, B,\ldots=1,\ldots ,6)$:
\be
g_{AB}=\bar g_{AB}+ \hat{\e} \, h_{AB}, \qquad C^{(2)}_{AB}=\bar C^{(2)}_{AB}+ \hat{\e} \, C_{AB} \,. 
\ee
The  equations for the perturbation are obtained by expanding to linear order the equations
\bea
R_{AB}&=&{1\over 4}F^{(3)}_{ACD}F^{(3)}_B{}^{CD}  \,, \nn
{}*_6 \! F^{(3)}&=&F^{(3)} \,, \qquad F^{(3)}_{ABC}=\p_A C^{(2)}_{BC}+\p_B C^{(2)}_{CA}+\p_C C^{(2)}_{AB} \,.
  \label{feq}
\eea
A $U(1)$ current $J^i_{-n}$ involves the torus direction $z_i$.  Let us choose the direction $z_1\equiv z$. This time we cannot consider just the 6D reduced theory -- the perturbation will involve the graviton $h_{Az}$ and the gauge field $C^{(2)}_{Az}$. To linear order in the perturbation, we still have $F^2=0$, so that $R=0$. The equations for the perturbation are obtained from linearizing about the background the 10D equations ($M,N\ldots=1, \dots ,10$)
\bea
R_{MN}&=&{1\over 4}F^{(3)}_{MPQ}F^{(3)}_N{}^{PQ} \,, \nn
F^{(3)}_{MNP}{}^{;P}&=&0, \qquad F^{(3)}_{MNP}=\p_M C^{(2)}_{NP}+\p_N C^{(2)}_{PM}+\p_P C^{(2)}_{MN} \,.
\label{feqq}
\eea

\subsection{Requirements of the perturbation}

We seek a perturbation which satisfies the following criteria: % set out in the introduction

\begin{enumerate}[(i)]
	\item The perturbation is a pure diffeomorphism in the throat background \eq{adsg}. The diffeomorphism at the upper end of the throat (boundary of $AdS$) should be generated by a vector field corresponding to one of $L_{-n}, J^a_{-n}, J^i_{-n}$.
	\item The perturbation should {\it not} be pure gauge in the neck, and should die away at infinity sufficiently rapidly to be normalizable.
	\item When we continue the solution down the throat, it should match onto a regular solution in the cap background  \eq{innerns}, as we describe in the next section.
	\item The perturbation should have a dependence $e^{-in(t-y)/R_y}$, so that it  has energy and momentum $E=P=n/R_y$. 
\end{enumerate}

Note that it is not obvious from the differential equations satisfied by the perturbation that such solutions will exist. We have chosen to make the perturbation go over to a diffeomorphism in the throat. With these requirement,  the solution can in general go over into a linear combination of growing and decaying solutions at spatial infinity. The fact that it goes over to only the decaying solution follows, of course, from the fact that there is indeed a state with $E=P$ for this physical system.

\subsection{The matching process}\label{secmatching}

The outer and inner regions overlap in the throat region
\be
a\ll r \ll \sqrt{Q} \,
\label{overlap}
\ee
whose metric is given in \eq{adsg}. We require the solutions in the outer and inner regions to agree in this overlap region; this will produce a solution that is regular everywhere. This matching process was developed in detail in \cite{Mathur:2003hj}; here we summarize the main steps.

A perturbation on the geometry (\ref{el}) satisfies a linear differential equation, which we write symbolically as $\t \square \phi=0$. Here $\phi$ denotes the perturbation, and $\t\square$ is a differential operator acting on the perturbation. Suppose we are solving the perturbation in the outer region. In the outer domain $r\gg a$ the full metric (\ref{el}) can be expanded as 
\be
g=g_{outer}+\epsilon g_1
\ee
where $g_0$ is the metric (\ref{outerr}) and $g_1$ captures the difference between the full metric $g$ and its approximation $g_{outer}$. The differential operator $\t\square$ can be correspondingly expanded as
\be
\t\square_g=\t\square_{g_{outer}}+\epsilon \square'
\ee
where $\t\square_{g_{outer}}$ is the wave equation for the perturbation in the leading order metric $g_{outer}$ and $\epsilon\square'$ captures all the subleading effects of the metric correction $g_1$. 

The perturbation itself can now be expanded as
\be
\phi=\phi_0+\epsilon \phi_1+\epsilon^2\phi_2+\dots
\ee
The equation satisfied by $\phi$ is
\be \label{eq:ep_exp}
\Big (\t\square_{g_{outer}}+\epsilon \square'_{outer}\Big ) \left(\phi^{outer}_0+\epsilon \phi_1^{outer}+\epsilon^2\phi_2^{outer}+\dots \right)= 0 \,.
\ee
At leading order in $\epsilon$ we get
\be
\t\square_{g_{outer}}\phi_0^{outer}=0
\ee
so we just have to solve the wave equation in the geometry (\ref{outerr}). At next order in $\epsilon$ we would have to solve
\be
\t\square_{g_{outer}}\phi_1^{outer}=-\square'_{outer}\phi_0^{outer}
\ee
which is the wave equation with source term $-\square'\phi_0^{outer}$. This process can be continued in a similar manner to all other orders in $\epsilon$. 

Now consider the inner region, and write the wave equation for $\phi^{inner}$ with a similar expansion to \eq{eq:ep_exp}.
The leading order solution satisfies 
\be
\t\square_{g_{inner}}\phi_0^{inner}=0
\ee
and we can continue to higher orders in epsilon as before.

Now suppose we have found a solution $\phi^{outer}_0$ in the outer region that dies off at $r\r \infty$, and a solution $\phi_0^{inner}$ that is regular in the cap.  For a complete solution to exist, it must be possible to choose a normalization for the outer and inner solutions such that 
\be
\phi_0^{outer}=\phi_0^{inner}+O(\epsilon), \qquad ~~a\ll r\ll \sqrt{Q}
\label{lowest}
\ee
where $O(\epsilon)$ denotes terms that vanish as $\epsilon\r 0$. 

This matching process can be carried out to arbitrarily many orders in $\epsilon$ (it was carried out to order $\e^3$ in \cite{Mathur:2003hj}), but in the present paper we restrict ourselves to noting agreement at the lowest level (\ref{lowest}).

\section{$L_{-n}$ perturbation} \label{sec:Ln}

\subsection{Virasoro algebra and asymptotic diffeomorphisms of $AdS_3$}

We start with a physical discussion of the asymptotic diffeomorphisms of $AdS_3$ introduced by Brown \& Henneaux \cite{Brown:1986nw}.
We consider the $AdS_3$ part of the geometry (\ref{adsg}), dropping the $S^3$ and $T^4$ parts for now, and  set $Q=1$, obtaining
\be
ds^2=-r^2 dt^2+r^2 dy^2+{dr^2\over r^2} \,.
\label{qttwo}
\ee
The dual CFT is parametrized by the coordinates $t,y$, and we can imagine it to live at a `boundary surface'  $r=R_{CFT}$.

Let us write $u=t+y, v=t-y$ and make an infinitesimal  diffeomorphism 
\be
v\r v+\hat\epsilon e^{-ik v} \,
\ee
where here and throughout the paper we use $\hat\epsilon \ll 1$ as a control of the smallness of the perturbation. Then the $AdS_3$ metric changes as $g_{\mu\nu}\r g_{\mu\nu}+\hat\epsilon h_{\mu\nu}$ with
\be
h_{uv}={i kr^2\over 2} e^{-i k v} \,.
\ee
In the CFT this corresponds to a `holomorphic diffeomorphism', which changes the 2D metric by a conformal factor.

In the CFT we can cancel this conformal metric change by a Weyl scaling $g_{ij}\r \Omega^2 g_{ij}$. In the dual $AdS_3$ geometry we can perform the analog of this rescaling by moving the location $r=R_{CFT}$ radially outwards or inwards by an infinitesimal amount, since the metric induced on the `boundary surface' scales  as $r^2$.  Thus we make a diffeomorphism $r\r r+\delta r$ such that
\be
\delta g_{uv}=-r\delta r = -\hat\epsilon h_{uv}=-\hat\epsilon {i kr^2\over 2} e^{-i k v}
\ee
which gives
\be
\delta r=\hat\epsilon \, {i kr\over 2}\,e^{-ikv} 
\ee
and now the metric on the surface $r=R_{CFT}$ is unchanged to leading order. We do however generate a subleading part $h_{vr}={k^2\over 2r}e^{- ik v}$. This can be cancelled by a shift $u\r u-\hat\epsilon { k^2\over 2r^2} e^{-ikv}$, leaving
\be
h_{vv}=-{ik^3\over 2}e^{-ikv} \,.
\label{qtone}
\ee
The perturbation (\ref{qtone}) is small enough at large $r$ to be counted among the allowed boundary diffeomorphisms that form the asymptotic symmetry group of $AdS_3$. (Measured in a local orthonormal frame, it falls as ${1\over r^2}$.)

For our purposes, we normalize the combination of the above diffeomorphisms as follows. The coordinate $y$ in our solution (\ref{outerr}) has period $2\pi R_y$, so we set $k={n\over R_y}$, with $n$ an integer. 
We require the vectors $\xi^L_n$ generating the $L_{-n}$ diffeomorphism to satisfy the commutation relations of the Virasoro algebra (without central charge)
\be
\left[ \xi^L_m, \xi^L_n \right] ~=~ i(m-n)\xi^L_{m+n} \,.
\ee
In order to do this we normalize the components of $\xi^L_n$ to be
\bea
\left( \xi^L_n \right)^v ~=~ R_y \,e^{-in \frac{v}{R_y}} \,, \quad 
\left( \xi^L_n \right)^r &=& \frac{i}{2} n r \, e^{-in\frac{v}{R_y}}\,, \quad
\left( \xi^L_n \right)^u ~=~ \frac{n^2}{2} \frac{Q^2}{r^2 R_y} \, e^{-in \frac{v}{R_y}} \,. \qquad \label{eq:xi_L_n} 
\eea
The metric perturbation (\ref{qtone}) then becomes 
\bea
h_{vv} &=&  -\frac{in^3}{2} \frac{Q}{R_y^2} \, e^{-in \frac{v}{R_y}} \,.
\label{hvv}
\eea
Since we are using an exponential $e^{-in \frac{v}{R_y}}$, we see an overall $i$ in the metric; to obtain the classical fields, one must take the real part of the perturbation. So a diffeomorphism with parameter $\cos{\frac{nv}{R_y}}$ produces a metric perturbation of the form $\sin{\frac{nv}{R_y}}$.

Note that the perturbation \eq{hvv} vanishes for the action of the $L_0$ diffeomorphism but not for $L_1$ and $L_{-1}$; this is because we are working in the Poincare patch and so are using the asymptotic form \eq{eq:xi_L_n} of the diffeomorphisms rather than the full expressions relevant for global $AdS_3$ (see e.g.~\cite{Maldacena:1998bw}).
This is also correct from the point of view of the CFT, where $L_{-1} \ket{\psi}$ should be non-zero for generic Ramond ground states $\ket{\psi}$.

\subsection{$L_{-n}$ perturbation in the outer region}

As it stands, (\ref{qtone}) is a just a diffeomorphism of $AdS_3$, and it is not clear in what sense it will create a new state $L_{-n} \ket{\psi}$. 
We next present a perturbation that equals the above diffeomorphism in the throat $r \ll \sqrt{Q}$, is {\it not} a pure diffeomorphism at the neck $r \sim \sqrt{Q}$, and which dies away at infinity. The energy carried by the nontrivial deformation in the neck region will account for the increase in energy expected from the application of an operator $L_{-n}$ to a CFT state. 

We construct this perturbation by taking the pure-diffeomorphism perturbation \eq{hvv} and making an ansatz in the full outer region metric (\ref{outerr}). The ansatz consists of multiplying the perturbation \eq{hvv} by an arbitrary smooth function of $r$. Solving the field equations (\ref{feq}) in the full outer region metric then yields the solution
\bea
h_{vv} &=& -\frac{in^3}{2} \frac{Q}{R_y^2} \, e^{-in \frac{v}{R_y}} \frac{Q}{Q+r^2} \,, \qquad C ~=~ 0 \,.
\label{qttw}
\eea
It can be seen that this perturbation is normalizable at infinity: the energy is proportional to $\,\int \! d^4 x (\p_r h_{vv})^2$ and so is finite. The perturbation is nontrivial in the neck $r\sim \sqrt{Q}$. 
In the throat $r \ll \sqrt{Q}$ must match onto a smooth solution in the cap, near $r=0$. In order to test this, we next study solutions of the field equations in the cap.

\subsection{Solution in the inner region} \label{sec:L_n_int}

In order to compare a perturbation around the outer background \eq{outerr} and a perturbation around the inner background \eq{innerr}, we must take account of the spectral flow transformation \eq{spectral}. In Appendix \ref{app:spectral} we show that in each of the perturbations we construct in this paper, this transformation gives only higher order corrections to the matching
in the throat, and thus has no effect at the leading order of matching that we  do.

The cap geometry we have chosen is global $AdS_{3}\times S^{3}\times T^{4}$, given in \eq{innerns}.
The equations for perturbations around $AdS_3\times S^3$ were found in \cite{Deger:1998nm}. The perturbation we are looking for does not involve the $T^4$, and will also turn out to be a singlet on the sphere. We take the ansatz for the metric perturbation
\be \label{eq:int_ansatz_h}
AdS_3:\quad h_{\mu\nu}=Mg_{\mu\nu}\,, \qquad\quad  S^3:\quad h_{ab}=Ng_{ab}
\ee
and for the perturbation of the $AdS$ field strength 
\be \label{eq:int_ansatz_F3}
F^{(3)}_{\mu\nu\lambda}=2(1+P)\epsilon_{\mu\nu\lambda} \,.
\ee
The gauge field on the sphere remains $F^{(3)}_{abc}=2\epsilon_{abc}$ since $\int_{S^3} F^{(3)}$ is a fixed number (the charge on the sphere). As shown in \cite{Deger:1998nm} and re-derived for completeness in Appendix \ref{app:L_n_int}, the equations of motion reduce to
\be \label{eq:sez_eqns}
M +3N =0 \,, \qquad P=2M \,, \qquad (\square-8)M=0 \,.
\ee
We seek a solution to $(\square-8)M=0$ with the ansatz $M=f(r) e^{-in{v\over R_y}}$ for some function $f(r)$. The solution which is regular at $r=0$ is
\bea \label{eq:sez_sol}
M&=& A \left(\frac{r^2}{r^2+a^2} \right)^\frac{n}{2} \left( n+1 + \frac{2r^2}{a^2} \right) e^{-in\frac{v}{R_y}} \,
\eea
for constant $A$.

\subsection{Matching in the throat region} \label{sec:L_n_match}

As discussed in Section \ref{secmatching}, the solutions in the outer and inner regions must match in the throat region,
\be
a ~ \ll ~ r ~ \ll \sqrt{Q} \,.
\ee
First consider the inner region solution given by \eq{eq:sez_eqns} and \eq{eq:sez_sol}. Taking the $r \gg a$ limit gives the leading order fields
\bea
M &=& 2 A \frac{r^2}{a^2} e^{-in \frac{v}{R_y}}\, , \qquad N ~=~ -\frac{1}{3} M \,, \qquad P ~=~ 2M \,.
\label{qwinner}
\eea
Next consider the outer region solution given by \eq{qttw}. We take $r \ll \sqrt{Q}$ and apply a diffeomorphism to 
match onto the above fields coming from the regular solution in the cap.  The diffeomorphism we use is generated by
\be \label{eq:Ln_undoing_diffeo}
\xi^v = - R_y e^{-i n \frac{v}{R_y}}  \, , \qquad 
\xi^r = -\frac{in}{2} e^{-in\frac{v}{R_y}}  \frac{Q^2 r}{(Q+r^2)^2} \,, \qquad
\xi^u = -\frac{n^2}{2} \frac{Q^2}{r^2 R_y} \, e^{-in \frac{v}{R_y}} \,.
\ee
Note that the diffeomorphism reduces to $-\xi^L_n$ in the $r \ll \sqrt{Q}$ limit (recall that $\xi^L_n$ generates the $L_{-n}$ diffeomorphism in the $AdS$ throat and is given in \eq{eq:xi_L_n}).
Applying this diffeomorphism to the outer region solution (\ref{qttw}) we get
\bea
h_{uv}&=&{3in\over 2}{r^4\over Q^2} e^{-in{v\over R_y}} \,, \qquad\quad h_{rr}~=~3in e^{-in{v\over R_y}} \,, \nn
h_{ab}&=&-in{r^2\over Q}e^{-in{v\over R_y}} \bar{g}_{ab} \,, \qquad C_{uv}~=~-{3in\over 2} {r^4\over Q^2} e^{-in{v\over R_y}}
\label{qefift}
\eea
where $\bar{g}_{ab}$ is the background $S^3$ metric, 
\be
\bar{g}_{ab} dx^a dx^b ~=~ Q \left[ d\th^2 + \cos^2 \th d \psi^2 + \sin^2 \th d \f^2 \right] .
\ee 
Using the definitions of $M, N, P$ in (\ref{eq:int_ansatz_h}), (\ref{eq:int_ansatz_F3}), we read off the fields
\bea
M &=& 3in \frac{r^2}{Q} e^{-in \frac{v}{R_y}} \, , \qquad N = -\frac{1}{3} M \,, \qquad P=2M \,.
\label{qwouter}
\eea
So we see that the inner region perturbation (\ref{qwinner}) and the outer region perturbation (\ref{qwouter}) agree if we choose
\be
A ~=~ {3in\over 2}\frac{a^2}{Q} ~=~ {3in\over 2} \frac{Q}{R_y^2} \,.
\ee

\subsection{Summary of $L_{-n}$ construction} \label{sec:Ln_summary}

Let us summarize what we have found above. Suppose we start with the particular 2-charge solution (\ref{el}), (\ref{selfdualfield}). This describes the D1D5 system in one of the Ramond ground states $|0\rangle_R$. Suppose we wish to derive the gravity solution that corresponds to the state $L_{-n}|0\rangle_R$. Then we add a perturbation that is given for $r\gg a$ by 
\bea
h_{vv} &=& -i n^3 \frac{Q}{R_y^2} \, e^{-in \frac{v}{R_y}} \frac{Q}{Q+r^2} \,, \qquad C ~=~ 0 \,
\label{qwfive}
\eea
and for $r\ll\sqrt{Q}$ by 
\be
M={3in\over 2}\frac{Q}{R_y^2}\left(\frac{r^2}{r^2+a^2} \right)^\frac{n}{2} \left( n+1 + \frac{2r^2}{a^2} \right) e^{-in\frac{v}{R_y}} 
\,, \quad N=-\frac{1}{3} M \,, \quad P=2M
\label{qwsix}
\ee
where $M,N,P$ are defined in (\ref{eq:int_ansatz_h}), (\ref{eq:int_ansatz_F3}). The expressions (\ref{qwfive}), (\ref{qwsix}) are gauge-equivalent in the overlap region $a\ll r\ll \sqrt{Q}$. Thus together they give a solution that is normalizable at infinity and regular in the cap.

At this stage it may appear that we can take the integer $n$ to be either positive or negative, but only the choice $n>0$ corresponds to a real excitation. This is because having found the linearized solution to the classical wave equation, we must quantize the perturbation to get the actual allowed excitations. In quantizing a scalar field we write
\be
\hat\phi={1\over (2\pi)^{3\over 2} \sqrt{2\omega}}[\hat a_{\vec k} e^{i\vec k\cdot \vec x-i\omega t}+\hat a_{\vec k} ^\dagger e^{-i\vec k\cdot \vec x+i\omega t}]
\ee
and only the function $e^{i\vec k\cdot \vec x-i\omega t}$ (with $\omega>0$) represents an allowed excitation. This is, of course, just a restatement of the fact that the vacuum has been selected to be the one that is annihilated by the $\hat a_{\vec k}$, so that it is the lowest energy state. In our case, the background solution $|0\rangle_R$ is a ground state for the given charges, so quantization of excitations will allow only positive energy excitations around it, which behave as $e^{-i\omega t}$ with $\omega>0$. This implies $n>0$ in (\ref{qwfive}). The energy of the quantized perturbation is $E=\omega$, which in our case is 
\be
E={n\over R_y} \,.
\ee
This is indeed the energy expected from an excitation $L_{-n}$ applied to a CFT state living on a circle of length $2\pi R_y$.

\section{Perturbation from sphere diffeomorphism $J^{(3)}_{-n}$} \label{sec:Jn}

\subsection{$SU(2)_L$ algebra on the sphere}

The CFT for the D1D5 system has a ${\cal N}=4$ supersymmetric chiral algebra in each of the left and right moving sectors. Thus besides the generators $L_{-n}$ we also have the bosonic generators  $J^a_{-n}$ as well as fermionic generators. We now turn to the diffeomorphisms corresponding to the $J^a_{-n}$ (we will not construct the perturbations for the fermionic generators in this paper). 

The $J^a_{-n}$ describe a current algebra, which arises from rotations of the sphere. We take the unit sphere $S^3$ to have the metric
\be
ds^2=d\theta^2+\cos^2\theta d\psi^2+\sin^2\theta d\phi^2 \,.
\ee
The symmetry group of this sphere is $SO(4)\approx SU(2)_L\times SU(2)_R$. The vector fields generating $SU(2)_L$ are
\bea \label{eq:SU2_gens}
V^{(3)}&=&\h(\p_\psi-\p_\phi) \,, \nn
V^{(+)}&=&\h e^{i(\psi-\phi)}\Big ( \p_\theta-i\tan\theta \p_\psi-i\cot\theta\p_\phi\Big ) \,, \\
V^{(-)}&=&\h e^{-i(\psi-\phi)}\Big ( \p_\theta +i\tan\theta \p_\psi+i\cot\theta \p_\phi\Big ) \nnm
\eea
with commutation relations
\be
[V^{(3)}, V^{(+)}]=i V^{(+)}, ~~~[V^{(3)}, V^{(-)}]=-i V^{(-)}, ~~~[V^{(+)}, V^{(-)}]=2i V^{(3)} \,.
\ee
From now on, we work with $J^{(3)}_{-n}$ because of the simpler form of $V^{(3)}$ compared to the other vector fields.
In particular $V^{(3)}$ generates rigid rotations of the $S^3$ by the diffeomorphism $\psi\r \psi +\h\hat\epsilon , ~~\phi\r\phi-\h\hat\epsilon$.

In order that the vectors $\xi^a_n$ generating the $J^a_{-n}$ diffeomorphism satisfy the appropriate commutation relations, i.e.
\bea
\left[ \xi^a_m, \xi^b_n \right] &=& i \e^{abc} \, \xi^c_{m+n} \,, \qquad 
\left[ \xi^L_m, \xi^a_n \right] ~=~ -i n \, \xi^a_{m+n} \,
\eea
we obtain $\xi^a_n = V^{(a)} e^{-in{v\over R_y}}$, which gives for $\xi^{(3)}_n$:
\be
(\xi^{(3)}_n)^\psi ~=~ \h e^{-in{v\over R_y}}, \qquad (\xi^{(3)}_n)^\phi=-\h e^{-in{v\over R_y}} \,.
\label{qeone}
\ee
This diffeomorphism applied to the  geometry of $AdS_3\times S^3$ generates a metric perturbation $h_{AB}$ but it also generates a perturbation of the gauge field $C^{(2)}$ since this gauge field is nonzero in the background solution. One can always accompany a diffeomorphism by a gauge transformation
\be
C^{(2)}_{AB}\r C^{(2)}_{AB}+\hat\epsilon \Big ( \p_A \Lambda_B-\p_B \Lambda_A\Big )
\label{gaugeq}
\ee
and we will next see that we indeed require such a gauge transformation in order to obtain an ansatz for the perturbation which is regular in the entire geometry (\ref{outerr}).

\subsection{Solution in the outer region}

We now consider the diffeomorphism generated by $\xi^{(3)}_n$, given in \eq{qeone}, applied to the throat $AdS_3\times S^3\times T^4$ geometry (\ref{adsg}). It generates the perturbation
\bea \label{eq:Ja_diff_fields}
h_{v \psi} &=& - \frac{in}{2} \frac{Q}{R_y} \cos^2\theta \, e^{-in \frac{v}{R_y}}  \,, 
\qquad C_{v \psi} ~=~ \phantom{-} \frac{in}{2} \frac{Q}{R_y} \cos^2\theta \, e^{-in \frac{v}{R_y}} \,, \cr
h_{v \phi} &=& \phantom{-} \frac{in}{2} \frac{Q}{R_y} \sin^2\theta \, e^{-in \frac{v}{R_y}}  \,, 
\qquad \, C_{v \phi} ~=~  \, \frac{in}{2} \frac{Q}{R_y} \cos^2\theta \, e^{-in \frac{v}{R_y}}  \,. \qquad
\eea
We will soon extend this to the full geometry (\ref{outerr}), but as it stands we have a difficulty: after introducing a dependence on $r$, the component $C_{v\phi}$ will lead to a field strength term $F^{(3)}_{v\phi r}$ which will have an angular dependence $\sim \cos^2\theta$. This is unacceptable, since in a unit orthonormal frame we will have $F^{(3)}_{\hat{r} \hat{v} \hat{\phi}} \sim \frac{\cos^2 \th}{\sin \th} $, which diverges at $\theta=0$. Thus we make a gauge transformation (\ref{gaugeq}) with 
\be \label{eq:Jn_lambda}
\L_\phi= Q \, e^{-in{v\over R_y}}
\ee
which yields
\bea \label{eq:Ja_diff_fields-2}
h_{v \psi} &=& - \frac{in}{2} \frac{Q}{R_y} \cos^2\theta \, e^{-in \frac{v}{R_y}}  \,, 
\qquad C_{v \psi} ~=~ \phantom{-} \frac{in}{2} \frac{Q}{R_y} \cos^2\theta \, e^{-in \frac{v}{R_y}} \,, \cr
h_{v \phi} &=& \phantom{-} \frac{in}{2} \frac{Q}{R_y} \sin^2\theta \, e^{-in \frac{v}{R_y}}  \,, 
\qquad \, C_{v \phi} ~=~  - \frac{in}{2} \frac{Q}{R_y} \sin^2\theta \, e^{-in \frac{v}{R_y}}  \,. \qquad
\eea
To extend this to a perturbation in the full outer region (\ref{outerr}), we take an ansatz where we multiply each of the above terms by a smooth function of $r$. The field equations (\ref{feq}) yield the solution
\bea
h_{v \psi} &=& - \frac{in}{2} \frac{Q}{R_y} \cos^2\theta \, e^{-in \frac{v}{R_y}} \frac{Q}{Q+r^2} \,, 
\quad C_{v \psi} ~=~ \frac{in}{2} \frac{Q}{R_y} \cos^2\theta \, e^{-in \frac{v}{R_y}} \frac{Q}{Q+r^2} \,, \cr
h_{v \phi} &=& \frac{in}{2} \frac{Q}{R_y} \sin^2\theta \, e^{-in \frac{v}{R_y}}  \frac{Q}{Q+r^2} \,, 
\quad \, C_{v \phi} ~=~ - \frac{in}{2} \frac{Q}{R_y} \sin^2\theta \, e^{-in \frac{v}{R_y}}  \frac{Q}{Q+r^2} \,. \qquad
\label{eq:sphere_outer}
\eea
This solution is normalizable at infinity, and nontrivial in the neck $r\sim \sqrt{Q}$. 
In the throat $r\ll \sqrt{Q}$ the solution matches to a solution in the inner region as we shall see in Section \ref{sec:Jn_match}.

\subsection{Solution in the inner region}

In the inner region $r \ll \sqrt{Q}$, we make the ansatz
\be
h_{\mu a}(x, y)=K_\mu(x) Y_a(y), \qquad C_{\mu a}(x, y)=Z_\mu(x) Y_a(y)
\label{ansatzaq}
\ee
where as before $\mu$ is an $AdS_3$ coordinate and $a$ is an $S^3$ coordinate. We must choose the spherical harmonic corresponding to the diffeomorphism (\ref{qeone}), i.e.
\be
Y_\theta=0 \,, \quad Y_\psi=\cos^2\theta \,, \quad Y_\phi=-\sin^2\theta \,.
\label{kk}
\ee
In Appendix \ref{sec:J_n_int} we study the field equations (\ref{feq}) under this ansatz and show that they yield the following solution which is regular in the cap. $Z_\mu$ is given by
\bea
Z_v &=&  C \, a \Big ( {r^2\over r^2+a^2}\Big ) ^{n\over 2} \left( n+1 + \frac{2r^2}{a^2} \right) e^{-i n {v\over R_y}}   \,,  \cr
Z_u &=&  C \, a \Big ( {r^2\over r^2+a^2}\Big ) ^{n\over 2} e^{-i n {v\over R_y}}    \,, \label{eq:sphere_inner} \\
Z_r &=&  C \, in {a^2 Q\over r(r^2+a^2)} \Big ( {r^2\over r^2+a^2}\Big ) ^{n\over 2} e^{-i n {v\over R_y}}    \, \nnm
\eea
for constant $C$, and $K_\mu$ is given by
\be
K_\mu ~=~ -2Z_\mu \,.
\ee
We next show that the perturbations in the inner and outer regions are gauge equivalent in the throat region.

\subsection{Matching in the throat region} \label{sec:Jn_match}

We now examine the solutions in the throat region
\be
a ~ \ll ~ r ~ \ll \sqrt{Q} \,.
\ee
In relating the inner and outer regions we have to use the spectral flow transformation \eq{spectral},
but as we show in Appendix \ref{app:spectral} the effect of this only arises at higher orders in perturbation theory.

We first take the $r \gg a$ limit in the inner region solution \eq{eq:sphere_inner}, which gives the leading order fields
\be
K_v=-4C{r^2\over a}e^{-in{v\over R_y}}, \qquad Z_v=2C{r^2\over a}e^{-in{v\over R_y}} \,.
\ee
Using (\ref{kk}) this corresponds to 
\bea 
h_{v \psi} &=&  -4C{r^2\over a} \cos^2\theta \, e^{-in \frac{v}{R_y}}  \,, 
\qquad C_{v \psi} ~=~  2C{r^2\over a} \cos^2\theta \, e^{-in \frac{v}{R_y}}  \,, \cr
h_{v \phi} &=& \phantom{-}4C{r^2\over a}  \sin^2\theta \, e^{-in \frac{v}{R_y}}  \,, 
\qquad  C_{v \phi} ~=~ -2C{r^2\over a}  \sin^2\theta \, e^{-in \frac{v}{R_y}} \,. \qquad \label{eq:h_Jn_ext_expandq}
\eea
We next take $r \ll \sqrt{Q}$ in the outer region solution \eq{eq:sphere_outer} and apply the diffeomorphism and gauge transformation generated by
\bea
\xi_n^{\psi} ~=~ -e^{-in \frac{v}{R_y}} \,, \qquad
\xi_n^{\phi} ~=~   e^{-in \frac{v}{R_y}} \, , \qquad  \L_\phi=-e^{-in{v\over R_y}}\label{eq:xi_J_nq} \,
\eea
which is the reverse of the $J^{(3)}_{-n}$ perturbation we started with, given in \eq{qeone} and \eq{eq:Jn_lambda}.
The resulting perturbation for $r\ll \sqrt{Q}$ has the form
\bea 
h_{v \psi} &=&  \phantom{-} i \frac{r^2}{R_y} \cos^2\theta \, e^{-in \frac{v}{R_y}}   \,, 
\qquad C_{v \psi} ~=~  - \, \frac{i}{2} \frac{r^2}{R_y} \cos^2\theta \, e^{-in \frac{v}{R_y}}   \,, \cr
h_{v \phi} &=& -  i \frac{r^2}{R_y} \cos^2\theta \, e^{-in \frac{v}{R_y}}    \,, 
\qquad  C_{v \phi} ~=~ \phantom{-} \frac{i}{2} \frac{r^2}{R_y} \cos^2\theta \, e^{-in \frac{v}{R_y}}  \,. \qquad \label{eq:h_Jn_ext_expandw}
\eea
We see that we get agreement between the inner region solution (\ref{eq:h_Jn_ext_expandq}) and the outer region solution (\ref{eq:h_Jn_ext_expandw}) if we choose
\be
C ~=~ \frac{i}{4} \frac{a}{R_y} ~=~ \frac{i}{4} \frac{Q}{R_y^2} \,.
\ee

\section{Perturbation from torus diffeomorphisms $J^i_{-n}$} \label{sec:Jz}

We next construct the perturbation arising from the $J^i_n$ diffeomorphisms, proceeding analogously to the previous sections. 
This perturbation is technically very similar to the perturbation from the sphere diffeomorphisms; the main difference between the two cases comes from the fact that $S^3$ has curvature, while $T^4$ does not. Since the $T^4$ is flat, the $z_i$ do not mix, and we can let the $z_i$ index lie along any one of the torus cycles, say $z_1\equiv z$.

The $J^i_{-n}$ do not have any non-trivial commutation relations amongst themselves, there is simply the commutation relation with the $L_{-n}$ generators. The vector $\xi_n^z$ generating the $J^z_{-n}$ diffeomorphism must then obey
\bea
\left[ \xi^L_m, \xi^z_n \right] &=& -i n \, \xi^z_{m+n} \,,
\eea
where $\xi^L_m$ is defined in \eq{eq:xi_L_n}, so we take $\xi_n^z$ to be
\be \label{eq:xi_Jz}
\xi_n^z ~=~ e^{-in{v\over R_y}} \,.
\ee

\subsection{Solution in the outer region}

Following the same procedure as before, we apply the diffeomorphism (\ref{eq:xi_Jz}) to the throat background (\ref{outerr}). This gives the perturbation
\be
h_{vz} ~=~ -in{R_z^2 \over R_y} e^{-in{v\over R_y}} \,.
\ee
As we did for the case of $J^{(3)}_{-n}$, we also perform a gauge transformation $C^{(2)}\r C^{(2)}+d\Lambda^{(1)}$, choosing here \be \label{eq:Lambda_Jz}
\Lambda^{(1)}_z ~=~ - R_z^2 \,e^{-in{v\over R_y}} \,.
\ee
This generates the field
\be
C_{vz} ~=~ in{R_z^2 \over R_y} e^{-in{v\over R_y}} \,.
\ee
To find a solution that is not pure gauge everywhere we assume an ansatz where we multiply the fields $h_{vz}, C_{vz}$ above by a smooth function of $r$. The field equations yield the solution 
\bea \label{eq:torus_outer}
h_{v z} &=& -in \frac{R_z^2}{R_y} \, e^{-in \frac{v}{R_y}} \frac{Q}{Q+r^2}  \,, 
\qquad C_{v z} ~=~  in \frac{R_z^2}{R_y} \, e^{-in \frac{v}{R_y}} \frac{Q}{Q+r^2}  \,.
\eea
As before, this solution is normalizable at infinity and in the throat $r\ll \sqrt{Q}$ it matches on to a smooth solution in the cap, as we show in Section \ref{sec:Jz_match}.

\subsection{Solution in the inner region} \label{sec:Jz_n_int}

In the inner region we use the following notation for the torus perturbation:
\be \label{eq:kz_z}
h_{\mu z} ~=~ K_\mu, \qquad C_{\mu z} ~=~ Z_\mu \,.
\ee
In Appendix \ref{appcq} we find that the field equations (\ref{feqq}) yield the solution
\bea
Z_v &=&  D \, a \Big ( {r^2\over r^2+a^2}\Big ) ^{n\over 2} \left( n+1 + \frac{2r^2}{a^2} \right) e^{-i n {v\over R_y}}   \,,  \cr
Z_u &=&  D \, a \Big ( {r^2\over r^2+a^2}\Big ) ^{n\over 2} e^{-i n {v\over R_y}}    \,, \label{eq:torus_inner} \\
Z_r &=&  D \, in {a^2 Q\over r(r^2+a^2)} \Big ( {r^2\over r^2+a^2}\Big ) ^{n\over 2} e^{-i n {v\over R_y}}  \, \nnm
\eea
for constant $D$, with $K$ given by 
\bea
K_\mu &=& -Z_\mu \,.
\eea
Note that the relation between $K$ and $Z$ is the main difference between the inner region solutions coming from the sphere diffeomorphism and those coming from the torus diffeomorphism. In the sphere case, we have $K_\mu = -2Z_\mu$. The difference can be traced to the fact that the sphere has curvature, unlike the torus.

\subsection{Matching in the throat region} \label{sec:Jz_match}

We now examine the solutions in the throat region
\be
a ~ \ll ~ r ~ \ll \sqrt{Q} \,.
\ee
We first take the $r \gg a$ limit in the inner region solution \eq{eq:torus_inner}, which gives the leading order fields
\be
K_v ~=~ -2D \,  {r^2\over a}e^{-in{v\over R_y}},
 ~\qquad Z_v ~=~ 2D \, {r^2\over a}e^{-in{v\over R_y}} \,.
\ee
Using (\ref{eq:kz_z}) this corresponds to 
\be 
h_{v z} ~=~  -2D \, {r^2\over a}  \, e^{-in \frac{v}{R_y}}  \,, 
\qquad C_{v z} ~=~  2D \,  {r^2\over a}  \, e^{-in \frac{v}{R_y}} \,.
\label{qefourt}
\ee
We next take $r \ll \sqrt{Q}$ in the outer region solution \eq{eq:torus_outer} and apply the diffeomorphism and gauge transformation generated by
\be
\xi^z ~=~ -e^{-in{v\over R_y}}, \qquad \Lambda^{(1)}_z ~=~ R_z^2 \, e^{-in{v\over R_y}}
\ee
which is the reverse of the $J^z_{-n}$ perturbation we started with, given in \eq{eq:xi_Jz}, \eq{eq:Lambda_Jz}.
The resulting perturbation for $r\ll \sqrt{Q}$ has the form
\bea
h_{vz} &=& in \frac{R_z^2}{R_y} \frac{r^2}{Q} \, e^{-in \frac{v}{R_y}}  \,, \qquad
C_{vz} ~=~ -in \frac{R_z^2}{R_y}  \frac{r^2}{Q} \, e^{-in \frac{v}{R_y}} \,.
\label{qetw_z}
\eea
We see that (\ref{qetw_z}) agrees with (\ref{qefourt}) if we choose
\be
D ~=~ - \frac{i}{2} \frac{a R_z^2}{R_y Q} ~=~ - \frac{i}{2} \frac{R_z^2}{R_y^2}  \,.
\ee

\section{A cross-check of our results} \label{sec:Check}

In the preceding sections we have found approximate perturbations around the background geometry \eq{el} and shown that the perturbations have the right properties to be dual to states created by generators of the chiral algebra of the D1-D5 CFT.

Our identification of states is based on~\cite{Brown:1986nw}. Let us ask if we can obtain a cross-check on this identification using general results from the AdS/CFT correspondence. Since the background geometry \eq{el} is asymptotically flat,
to carry out such a check we must decouple the $AdS$ region. 
A test will then be provided by calculating the one-point functions of chiral primaries from the gravity and CFT sides~\cite{deHaro:2000xn}. Such tests have been done for the spectral flow coordinate transformation~\cite{Hansen:2006wu,*Kraus:2006nb} and for general two-charge D1-D5 geometries~\cite{Skenderis:2006ah,*Kanitscheider:2006zf}.

In this section we carry out such a  test, and exhibit the required agreements. Thus our construction based on~\cite{Brown:1986nw} is consistent with the above-mentioned $AdS$/CFT work.
We follow in places~\cite{Hansen:2006wu,*Kraus:2006nb,Skenderis:2006ah,*Kanitscheider:2006zf}.
%, following most closely~\cite{Skenderis:2006ah,*Kanitscheider:2006zf}.

\subsection{The $L_{-n}$ perturbation}

We first derive the one-point function for the stress-energy tensor in the $L_{-n}$ perturbation.
On the CFT side, consider the state
\be
\ket{\mu} ~=~ \left( \one + \mu L_{-n} \right) |0\rangle_R \,
\ee
where we have introduced the (complex) parameter $\mu$. By the standard relation between  classical perturbations and coherent states,  $\mu$ will be  proportional to $\hat\epsilon$, the parameter controlling the size of the classical perturbation;  thus we work to linear order in $\mu$.
The above state is proposed to be dual to the perturbed metric constructed in Section~\ref{sec:Ln}. 

In this state, to linear order in $\mu$ the only nonzero expectation values are those for the operators $L_{\pm n}$. The mode $L_{n}$ has expectation value
\bea \label{eq:Ln_cft_vev}
\bra{\mu} L_n \ket{\mu} ~=~ \mu \frac{n_1 n_5}{2} n^3 \,
\eea
where we have used 
\bea 
\com{L_m}{L_n} 
  &=& \frac{c}{12} (m^3-m) \delta_{m+n, 0} + (m-n) L_{m+n} \,, \qquad c = 6 n_1 n_5 \,.
\eea
The mode $L_{-n}$ has an analogous one-point function; since have written our classical perturbation in the form $e^{-in \frac{v}{R_y}}$ (and not $\cos (n \frac{v}{R_y})$) we shall deal only with $L_{n}$. The energy $L_0$ of the perturbation appears at quadratic order in $\mu$, and so is beyond the scope of this paper.

On the gravity side, the constant mode at the asymptotically $AdS$ part of the inner region is the part generated by the $L_{-n}$ diffeomorphism, which from \eq{hvv} is
\bea
h_{vv} &=&  -\frac{in^3}{2} \frac{Q}{R_y^2} \, e^{-in \frac{v}{R_y}} \,.
\label{hvv-2}
\eea
To compare to the CFT stress tensor, we make the Fefferman-Graham expansion using the coordinate $z=Q/r$. We temporarily use $\m, \n$ for the remaining two $AdS_3$ directions. The expansion is
\bea
ds^2 ~=~ \frac{Q}{z^2} \left[ dz^2 +  \Big( g_{(0)\m\n} + z^2  g_{(2)\m\n} + \cdots 
 \! \Big) \, dx^{\m} dx^{\n} \right] \,, 
\eea
where `$\cdots$' indicates terms which do not contribute to our checks. Using complex coordinates\footnote{We define the coordinates $\sigma = \frac{y}{R_y} \,,  \tau = \frac{t}{R_y}$
so that $\sigma$ has period $2\pi$. We identify the CFT base space coordinates as $\sigma, \tau$, and
in terms of the Euclidean time $\tau_E = i \tau$, we define $w = \sigma + i \tau_E  = \sigma - \tau .$
}, the one-point function for the holomorphic stress-energy tensor is given at linear order in $\hat\epsilon$ by~(see e.g.~\cite{Skenderis:2006ah,*Kanitscheider:2006zf})
\bea
\langle T_{ww} \rangle &=& n_1 n_5 \Big( g_{(2)ww} + \cdots \Big) \cr
 &=&  - \hat\epsilon \, \frac{in^3}{2} (n_1n_5) \, e^{in w} \,.
\eea
This implies that the mode $L_{n}\,$ has the one-point function
\be
\langle L_{n} \rangle ~~=~~ \frac{1}{2\pi} \oint dw \, e^{-inw} \langle T_{ww} \rangle 
~~=~~ - \hat\epsilon \, i \, n^3 \, \frac{n_1n_5}{2} \,.
\ee
Since $\mu$ is proportional to $\hat\epsilon$, this is consistent with the CFT one-point function \eq{eq:Ln_cft_vev}.

%
%To make the cross-check we then compare
%\bea
%\mu \, n^3  \frac{n_1 n_5}{2}  \quad \longleftrightarrow \quad \hat\epsilon \, i \, n^3 \, \frac{n_1 n_5}{2}.
%\eea
%We have argued that $\mu$ is proportional to $\hat\epsilon$ so $\mu = i \hat\epsilon$ is very reasonable (both numbers are complex).
%We could in principle attempt to cross-check also the coefficient of proportionality between $\mu$ and $\hat\epsilon$ 
%against an independent calculation; we expect that this may be possible at nonlinear order in $\hat\epsilon$. One could imagine normalizing the gravity perturbation, and identifying a normalized field mode with a single quantum created by $L_{-n}$. We leave this for future work.

\subsection{The $J^{(3)}_{-n}$ and $J^{z}_{-n}$ perturbations}

We now turn to the $J^{(3)}_{-n}$ and $J^{z}_{-n}$ perturbations, and for concreteness we work with $J^{(3)}_{-n}$. 
At the asymptotically $AdS$ part of the inner region, both the $J^{(3)}_{-n}$ and $J^{z}_{-n}$ perturbations have constant terms of the form $h_{AB}=-C_{AB}$. As a result, we expect all the steps to go through also for the $J^{z}_{-n}$ perturbation.

On the CFT side, consider the state 
\be
\ket{\lambda} ~=~ \left( \one + \lambda J^{(3)}_{-n} \right) |0\rangle_R \,.
\ee
In this state, to linear order in $\lambda$ the only nonzero expectation values are those for
the operators $J^{(3)}_{\pm n}$. Again we deal only with $J^{(3)}_{n}$ which has expectation value
\bea \label{eq:Jn_cft_vev}
\bra{\lambda} J^{(3)}_n \ket{\lambda} ~=~ \lambda \frac{n_1 n_5}{2} n \,
\eea
where we have used 
\bea
\com{J^a_m}{J^b_n} 
  &=& \frac{c}{12}m\delta^{ab}\delta_{m+n, 0} +i{\epsilon^{ab}}_cJ^c_{m+n} \,, \qquad c = 6 n_1 n_5 \,.
\eea
On the gravity side, we set up some notation for reading off the one-point functions.
Following our conventions for the $SU(2)$ generators \eq{eq:SU2_gens}, we form the combinations
\bea
g_{\mu}^{\,(3)} &=& \ha \left( g_{\mu}^{~\psi} - g_{\mu}^{~\phi} \right) \,, \qquad 
g_{\mu}^{\,\bar{(3)}} ~=~ \ha \left( g_{\mu}^{~\psi} + g_{\mu}^{~\phi} \right)   
\eea
and similarly for $C^{(2)}$. We then define the one-forms $A^{(3)}$, $A^{\bar{(3)}}$, by
\be
A^{(3)}_\mu ~=~  - \Big[ g_{\mu}^{\,(3)} - (C^{(2)})_{\mu}^{~(3)} \Big] \,, \qquad A^{\bar{(3)}}_\mu ~=~ 
- \left[ g_{\mu}^{\,\bar{(3)}} + (C^{(2)})_{\mu}^{~\bar{(3)}} \right] \,.
\ee
We now focus on the holomorphic $SU(2)_L$; analogous expressions apply for $SU(2)_R$. 
We then make the Fefferman-Graham expansion 
\be \label{eq:A_Feff}
A^{(3)} ~=~ \cA^{(3)} + z^2 A^{(3)}_{(2)} + \cdots
\ee
The one-point function for $J^{(3)}$ is then given by (see e.g.~\cite{Skenderis:2006ah,*Kanitscheider:2006zf})
\be \label{eq:J-A}
\langle J^{(3)} \rangle ~=~ \frac{n_1n_5}{2} \cA^{(3)} \,.
\ee
For the case at hand the constant mode at the asymptotically $AdS$ part of the inner region is the part generated by the $J^{(3)}_{-n}$ diffeomorphism. From \eq{eq:Ja_diff_fields-2},
at order $\hat\epsilon$ we obtain\footnote{Rather than introduce additional notation, we leave it as understood that the parameter $\hat\epsilon$ may not in general be the same between the different perturbations. }
\be
\cA^{(3)} ~=~ - \hat\epsilon \, i n \, e^{in w} dw   \qquad \Rightarrow \qquad 
\langle J^{(3)} \rangle ~=~ - \hat\epsilon \, i n \frac{n_1n_5}{2} \, e^{inw} dw \,.
\ee
This implies that the mode $J^{(3)}_{n}\,$ has the one-point function
\be
\langle J^{(3)}_{n} \rangle ~~=~~ \frac{1}{2\pi} \oint dw \, e^{-inw} \langle J^{(3)}_{w} \rangle 
~~=~~ - \hat\epsilon \, i n \frac{n_1n_5}{2} \,.
\ee
Since $\lambda$ is proportional to $\hat\epsilon$, this is consistent with the CFT one-point function \eq{eq:Jn_cft_vev}.

%\newpage

\section{Discussion} \label{sec:Discussion}

\subsection{Summary of our results}

Let us summarize our construction. We have wrapped $n_5$ D5 branes on $T^4\times S^1$ and $n_1$ D1 branes on $S^1$. Their bound state has $\exp[2\sqrt{2}\pi\sqrt{n_1n_5}]$ degenerate ground states $|\psi_i\rangle_R$, where the subscript $R$ denotes the Ramond sector.
We take the $S^1$ radius to be large compared to the $AdS$ curvature length scale;
then the low energy dynamics of the D1D5 bound state gives a 1+1 dimensional CFT along $t, y$, where $y$ parametrizes the $S^1$.  

The generic D1D5 bound state is quantum in nature and the gravitational field it sources is not a priori well-described by a classical geometry all the way into the cap region. Certain states do however have backreactions well-described by classical geometries (for a discussion see e.g.~\cite{Lunin:2002bj}); let $|\psi\rangle_R$ be such a state. 
Previous work has taught us that these geometries have the structure of flat space at infinity, then an intermediate `neck' region, 
then a `throat' which is approximately the Poincare patch of $AdS_3 \times S^3 \times T^4$, and a cap whose structure depends on the state $|\psi\rangle_R$.

Since we can describe the D1D5 system as a CFT, we can ask for the gravity description of a state such as $L_{-n}|\psi\rangle_R$, which carries D1, D5 and P charge.
Following the general ideas of \cite{Brown:1986nw} we expect that in this $AdS$ region the $L_{-n}$ operators are represented by diffeomorphisms that leave the throat asymptotically $AdS_3 \times S^3 \times T^4$. 
We can of course continue this diffeomorphism out past the neck in any way we like, but a pure diffeomorphism of the starting geometry will not give the gravitational solution created by $L_{-n}|\psi\rangle_R$, since the required solution must have more energy and momentum than the solution for the state $|\psi\rangle_R$. 

Taking an ansatz $e^{-in{(t-y)\over R_y}}$ to account for the energy and momentum, we find solutions that reduce to diffeomorphisms in the throat, but are {\it not} pure gauge in the neck. 
Our solutions are normalizable at infinity, and we match them to regular solutions in the particular cap we choose, namely the cap for the state obtained from spectral flow of the NS vacuum.

Proceeding in this manner, we obtain perturbations corresponding to the generators of the chiral algebra $L_{-n}, J^{(3)}_{-n}$ in the D1D5 CFT, as well as for the four chiral $U(1)$ currents $J^i_{-n}$ arising from translation symmetry along the four directions of $T^4$. As noted in Section \ref{sec:Ln_summary}, upon quantization only the perturbations with $n>0$ correspond to physical excitations.

Since the nontrivial part of the perturbation is concentrated at the neck which forms a natural boundary to the $AdS$ region, it is plausible that the states we find are in some way related to the `singleton' (or `doubleton') representations that lie at the boundary of $AdS$ \cite{Dirac:1963ta,*Flato:1978qz,*Gunaydin:1986fe}, see also~\cite{deBoer:1998ip}.\footnote{In \cite{Maldacena:2001ss} an approximate computation of a singleton boundary mode was given for the NS5 brane geometry.}  Note however that the neck is not part of $AdS$ itself, and it breaks the rotational $SO(4)$ symmetry of the $S^3$. This happens because the physical D1D5 states are in the Ramond (R) sector, while the global $AdS_3\times S^3$ geometry describes the NS vacuum. The relation between R and NS sector states is given by a `spectral flow' (\ref{spectral}) which uses a choice of direction in the $S^3$.

A consequence of this breaking of symmetry is that while $L_0,L_1, L_{-1}$ are isometries of global $AdS_3\times S^3$, they differ in their effect on the physical $R$ sector state given by the solution (\ref{el}). $L_0$ is still a symmetry of the $R$ sector solution, so it generates no perturbation, but $L_{-1}$ is {\it not} a symmetry and does generate a nontrivial perturbation; this is of course in agreement with the fact that $L_{-1}|\psi\rangle_R\ne 0$ on R sector states. ($L_1$ is a annihilation operator that kills the $L_{-1}$ perturbation after quantization).

\subsection{Relation to previous attempts to construct `hair'}

In this paper we have considered linear perturbations, with the metric and gauge field given by $g_{AB}=\bar g_{AB}+\hat\epsilon h_{AB}, C^{(2)}_{AB}=\bar C^{(2)}_{AB}+\hat\epsilon C_{AB}$. It turns out however that in the outer region we can set $\hat\epsilon$ to be arbitrary, and obtain an exact solution of the field equations. These nonlinear solutions correspond to CFT operators like $\exp[\hat\epsilon L_{-n}]$ where $\hat\epsilon$ is no longer infinitesimal. 

In the case of the $U(1)$ currents in the torus directions, the nonlinear solutions in the outer region were previously found and studied in in~\cite{Larsen:1995ss,*Cvetic:1995bj,*Horowitz:1996th,*Horowitz:1996cj} for deformations of the black string. These solutions were, later found to be singular at the horizon~\cite{Kaloper:1996hr,*Horowitz:1997si}. Such a singularity can be expected because of the general idea of the `no hair theorem', namely that it is difficult to have any smooth perturbation of a horizon.

On a similar note, in \cite{Banerjee:2009uk} the extremal D1D5P black hole geometry (with horizon) was considered, and linearized perturbations around this geometry were constructed with support in the neck. 
In a following paper \cite{Jatkar:2009yd} it was found that the perturbations were singular at nonlinear order at the horizon, for the same reasons found in~\cite{Kaloper:1996hr,*Horowitz:1997si}. 

In our construction, however, there is no such singularity; the background geometry does allow the perturbations we construct.
It is interesting to note the differences between the solutions of \cite{Larsen:1995ss,*Cvetic:1995bj,*Horowitz:1996th,*Horowitz:1996cj,Banerjee:2009uk} and those of this paper.

\begin{enumerate}
	\item The geometries of \cite{Larsen:1995ss,*Cvetic:1995bj,*Horowitz:1996th,*Horowitz:1996cj,Banerjee:2009uk} had horizons, while we have a `cap'; thus our perturbations do not suffer from this singularities mentioned above. However we are required to impose regularity in the cap, which we do.\footnote{We examine only linear perturbations of the geometry; this is adequate since a regular perturbation on a smooth solution is not expected to be singular at nonlinear order.}
	\item For the purposes of \cite{Banerjee:2009uk} the perturbations were not required to exist as states, since they were only used to relate the state counts of the 5D black hole and the 4D black hole. They turn out to be singular at the horizon in both cases, and so drop out of both sides of the state count relation. In our case the perturbations correspond to states like $L_{-n}|\psi\rangle_R$, so it would be strange if they did not exist for the full asymptotically flat D1D5 geometry created by the branes.
	\item On a technical note, the perturbations in the $R^4$ directions in~\cite{Larsen:1995ss,*Cvetic:1995bj,*Horowitz:1996th,*Horowitz:1996cj,Banerjee:2009uk} are in an $l=1$ spherical harmonic while the closest analog for our case, the $L_{-n}$ perturbations, are in the $l=0$ harmonic.
\end{enumerate}

We note in passing that  in \cite{Banerjee:2009uk,Jatkar:2009yd} the term `hair' was used for  degrees of freedom living far away from an event horizon. This appears different, however, from the traditional terminology of `hair' used by relativists, where `hair' is used for perturbations {\it at} the  horizon.  
In this sense \cite{Larsen:1995ss,*Cvetic:1995bj,*Horowitz:1996th,*Horowitz:1996cj,Kaloper:1996hr,*Horowitz:1997si} were attempts to construct `hair',
while in \cite{Banerjee:2009uk,Jatkar:2009yd} and this paper there is a long AdS throat separating the perturbation from the cap or horizon. 
The long AdS throat is necessary so that one may apply the tools of AdS/CFT, and its length
is of order $\log \frac{1}{\e}$ where $\e = {\sqrt{Q}\over R_y}$, 
as discussed in Section \ref{sec:background_fields}.

\subsection{Future directions}

The boundary diffeomorphisms correspond to the application of the chiral algebra generators $L_{-n}, J^{(3)}_{-n}, J^z_{-n}$ to states $|\psi_i\rangle_R$. The space of states we can generate this way corresponds to a central charge $c=6$; this is of course in accord with the fact that in $AdS/CFT$ duality a $U(1)$ center of mass degree of freedom lives at the boundary of $AdS$ while the remaining states (corresponding to  $c=6(n_1n_5-1)$)  lie near $r=0$. In this sense the construction of this paper directly address only a very small family of microstates. But the construction also shows us a way of understanding states that do {\it not} live at the boundary, as we now discuss.

\begin{figure}[htbp]
\begin{center}
\includegraphics[width=15cm]{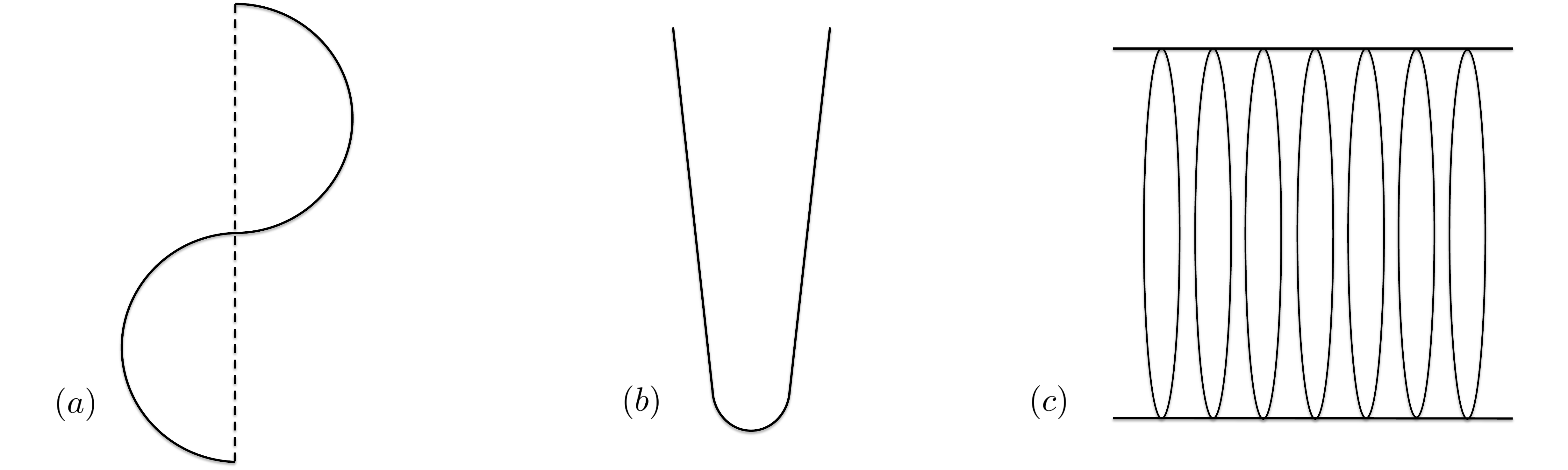}
\caption{The background geometry and corresponding CFT state used in this paper.}
\label{fig:4}
\end{center}
\end{figure}

The geometries for the two-charge extremal D1D5 states $|\psi_i\rangle_R$  are generated by taking a vibration profile for the string in the NS1P duality frame, finding the metric for this profile, and then dualizing to the D1D5 frame. When the vibration profile was a single turn of a helix (\ref{eq:helix_profile}), depicted in Fig.\;\ref{fig:4}\,$(a)$, we obtained global $AdS_3\times S^3$ for the D1D5 geometry \ref{fig:4}\,$(b)$. The corresponding D1D5 state $|0\rangle_R$ is described in the orbifold CFT~\cite{Seiberg:1999xz,Larsen:1999uk}
by the completely untwisted sector, depicted in Fig.\;\ref{fig:4}\,$(c)$. On this state we have
\be
L_{-n}|0\rangle_R=\sum_{k=1}^{n_1n_5}L^{(k)}_{-n}~\left(\prod_k |0_k\rangle_R \right)
\ee
where the index $k$ labels the $n_1n_5$ untwisted copies of the $c=6$ CFT which make up the state $|0\rangle_R$, and $L^{(k)}_{-n}$ is the operator $L_{-n}$ acting on the $k$th copy. Thus we get an action of the {\it diagonal} $L_{-n}$ arising from the set of individual $L^{(k)}_{-n}$.

\begin{figure}[htbp]
\begin{center}
\includegraphics[width=15cm]{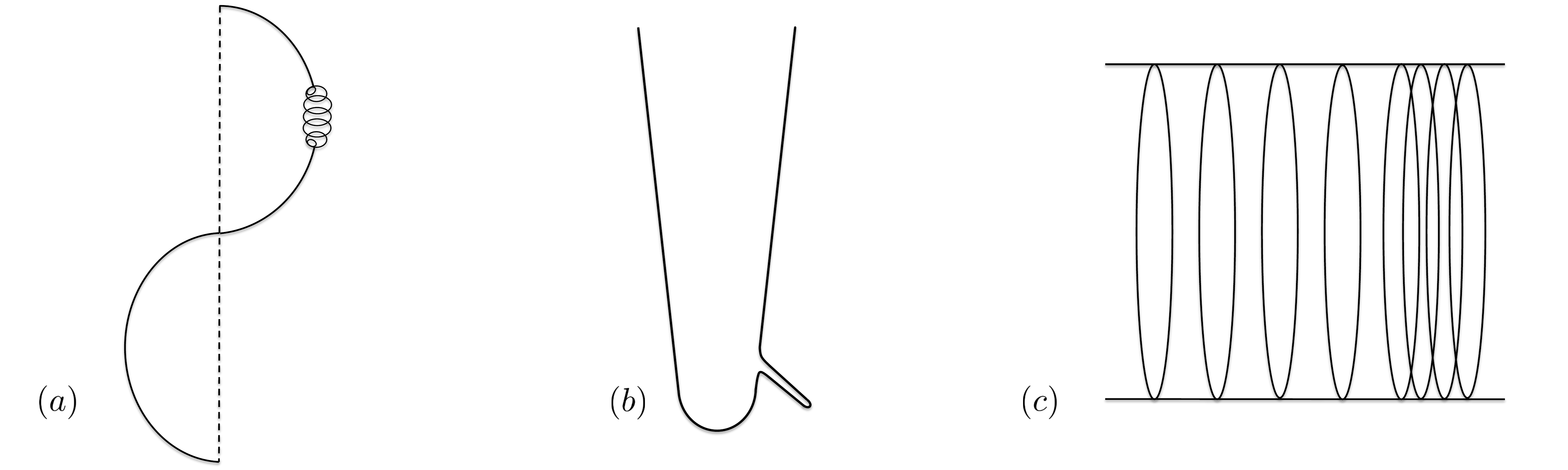}
\caption{Branching throat with one small subthroat.}
\label{fig:5}
\end{center}
\end{figure}

Now consider a state $|\psi\rangle _R$ generated by the NS1P profile in Fig.\;\ref{fig:5}\,$(a)$. The corresponding D1D5 geometry \ref{fig:5}\,$(b)$ has a small `subthroat' corresponding to the high frequency part of the NS1P profile, and in the orbifold CFT we obtain a state with a set of twisted strings \ref{fig:5}\,$(c)$. Since the subthroat is small, it opens into a part of the global $AdS$ that is close to flat space on the scale of the subthroat. Then the computations done in this paper tell us that in a suitable approximation one may construct modes localized at the neck of the subthroat. It is natural to identify  an $L_{-n}$ perturbation of the subthroat with the state
\be
|0_1\rangle_R\otimes |0_2\rangle_R\dots |0_n\rangle_R\otimes \sum_{k=1}^K L_{-n}^{(k)} \left(\prod_{k=1}^K |0_k^s\rangle_R \right)
\ee
where now the $L_{-n}$ operators have been applied to the set of twisted copies $|0^s_k\rangle_R$ ($s$ denotes the order of the twist), and $k=1, \dots K$ runs over the different  twisted copies.
 
Thus we see that the perturbations localized at the neck leading to flat space correspond to the diagonal action of $L_{-n}$ on all copies of the CFT, while similar modes located inside the `cap' region correspond to applying $L_{-n}$ {\it unequally} to the different copies of the CFT.

\begin{figure}[htbp]
\begin{center}
\includegraphics[width=15cm]{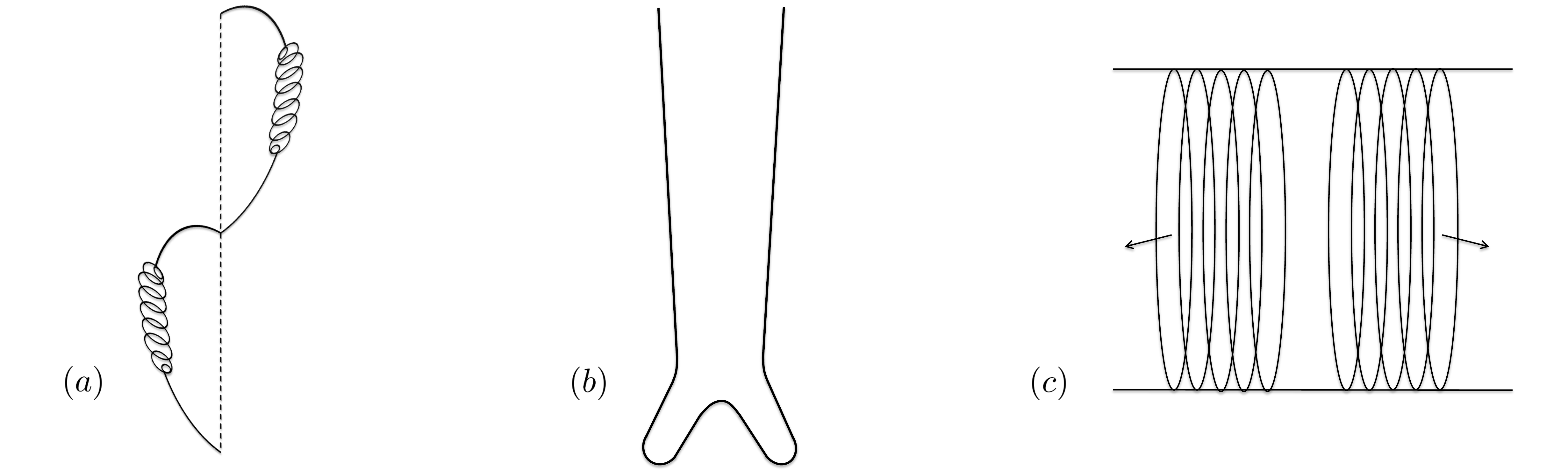}
\caption{Branching throat with two large subthroats.}
\label{fig:6}
\end{center}
\end{figure}

As another example, consider the branching throat geometry with two large subthroats depicted in Fig.~\ref{fig:6}\,$(b)$. For each subthroat we can look for a perturbation of the kind we have constructed, and we then look at the antisymmetric combination of these perturbations. This combination will cancel out as we go far above the junction region of the throats, and so the perturbation will be localized around this junction region itself. The corresponding CFT state would have the structure
\be
\left(\sum_{k=1}^{n_1n_5\over 2}L^{(k)}_{-n}~-~\sum_{k={n_1n_5\over 2}+1}^{n_1n_5}L^{(k)}_{-n}\right) 
\left(\prod_{k=1}^{n_1n_5\over 2}|0_k^+\rangle_R \otimes \prod_{k={n_1n_5\over 2}+1}^{n_1n_5}|0_k^-\rangle_R \right)
\ee
where the 2-charge extremal state corresponding to the branched-throat geometry has half the copies of the $c=6$ CFT in one spin state $|0_k^+\rangle_R$ and half in the opposite spin state $|0_k^-\rangle_R \;\!$. We see that the action of $L_{-n}$ is not the diagonal one that would give a mode localized at the neck leading to asymptotic infinity, but one where we apply the $L_{-n}$ with different signs to the two sets of CFT copies.
 
We hope to return for a more detailed analysis of such states elsewhere, but for now we just note that such states at the cap are examples of the `hair' which should replace the naive horizon geometry of the traditional black hole in the fuzzball picture.
In this picture there are no states with a regular horizon, and the above examples illustrate how some specific classes of CFT states may look in the gravity picture.

It would also be interesting to explore the types of perturbations considered in this paper in other setups in which there is a black hole with a locally $AdS_3$ throat, such as the extremal, vanishing horizon (EVH) black holes that have received interest recently \cite{SheikhJabbaria:2011gc,*deBoer:2011zt}, and to investigate relations to the elliptic genus of $AdS_3$ \cite{deBoer:1998us,Murthy:2011dk}. 
One could also attempt to construct similar perturbations in $AdS$ spacetimes of different dimensions; while there is a full Virasoro algebra only for $AdS_3$, we expect there to be perturbations corresponding to symmetry generators of the global symmetry groups of $AdS$ spacetimes of different dimensions.

An obvious further avenue is to carry out the construction in the present paper to next order in perturbation theory, and to investigate whether this can be generalized to an all-order solution. We plan to return to this in a future paper.

\section*{Acknowledgements}

We thank Dileep Jatkar and Yogesh Srivastava for collaboration at the early stages of this project. We thank Steve Carlip, Borun Chowdhury, Atish Dabholkar, Stefano Giusto, Murat Gunaydin, Vishnu Jejjala, Rodolfo Russo and  Ashoke Sen 
for helpful discussions and comments. 
We also thank the anonymous referee for comments which led to improvements of the paper.
This work was supported in part by DOE grant DE-FG02-91ER-40690.

\begin{appendix}

\section{Spectral flow} \label{app:spectral}

We have split the full geometry (\ref{el}) into two parts: an outer region (\ref{outerr}) and an inner region (\ref{innerr}). The angular coordinates on the $S^3$ are related by the spectral flow~\cite{Schwimmer:1986mf} map given in (\ref{spectral}),
\be
\psi_{NS} ~=~ \psi-{a\over Q}y, \quad \phi_{NS} ~=~ \phi-{a\over Q}t \,.
\label{spectralq}
\ee
We solve the wave equation in the outer region, then in the inner region, and then match the two solutions. For this matching, we should take into account the above transformation relating the two coordinates. But as we will now see, to the order where we are working, this change is {\it not} significant. If we wished to work to higher orders in the small parameter $\epsilon$ defined in (\ref{limitq}) then we would have to take (\ref{spectralq}) into account; this was done for example in \cite{Mathur:2003hj} to several orders in $\e$.

Suppose a  vector $V^{inner}_A$ is given in the inner region coordinates, and we  wish to find $V^{outer}_A$, the components in the outer region coordinates.  The relevant components change as follows:
\bea
A^{outer}_y ~=~ A^{inner}_y-{a\over Q}A^{inner}_{\psi_{NS}} \,, && \quad A^{outer}_\psi~ ~=~ ~A^{inner}_{\psi_{NS}} \,, \nn
A^{outer}_t ~=~ A^{inner}_t-{a\over Q}A^{inner}_{\phi_{NS}} \,, && \quad A^{outer}_\phi~ ~=~ ~A^{inner}_{\phi_{NS}} \,.
\eea
The changes $-{a\over Q}A^{inner}_{\psi_{NS}}, -{a\over Q}A^{inner}_{\phi_{NS}}$ must now be compared to the terms that did not change. To do this we go to a unit orthonormal frame, where we denote components by writing a `hat' over their index. We match the inner and outer region solutions in the throat $a\ll r\ll \sqrt{Q}$, so we use the metric (\ref{adsg}). The hatted expressions are then
\be
A^{outer}_{\hat\psi}~\sim~ {1\over \sqrt{Q}}A^{inner}_{\psi_{NS}},\qquad
A^{outer}_{\hat y}~\sim~ {\sqrt{Q}\over r}(A^{inner}_y-{a\over Q}A^{inner}_{\psi_{NS}}) \,.
\ee
We see that the change due to spectral flow in $A^{outer}_{\hat y}$  has magnitude
\be
\Delta A^{outer}_{\hat y}~\sim~ {a\over \sqrt{Q}r}A^{inner}_{\psi_{NS}}~\sim~ {a\over r}A^{outer}_{\hat\psi} \,.
\ee
Assuming a general scale for all components of the perturbation $A^{outer}_{\hat y}\sim A^{outer}_{\hat\psi}\sim \hat A$, we see that
\be
{\Delta A^{outer}_{\hat y}\over A^{outer}_{\hat y}} ~\sim~ {a\over r}~\ll~1 \,.
\ee
Thus the changes due to spectral flow are a higher order correction to the matching in the throat, and need not be considered at the leading order of matching that we perform.

\section{The $L_{-n}$ perturbation in the interior} \label{app:L_n_int}

The cap geometry we have chosen is global $AdS_{3}\times S^{3}\times T^{4}$. The solution for this geometry can be scaled to take the form
\bea
ds'^2&=&-(1+r'^2)dt'^2+{dr'^2\over 1+r'^2}+r'^2 dy'^2 + d\Omega_3^2+dz'_i dz'_i \,, \nn
F'^{(3)}_{abc}&=&2\epsilon_{abc} \,, \quad F^{(3)}_{\mu\nu\lambda} ~=~ 2\epsilon_{\mu\nu\lambda} \,,
\label{aone}
\eea
where
\be \label{eq:primed_coords}
ds'^2={ds^2\over Q}, ~~~t'={at\over Q}, ~~~y'={ay\over Q}, ~~~r'={r\over a}, ~~~z'_i=z_i, ~~~F'^{(3)}={F^{(3)}\over Q} \,.
\ee
We will drop the primes on the variables in what follows, and restore them at the end so that we can change back to the unprimed variables. 

The equations for small perturbations around $AdS_3\times S^3$ were written down in \cite{Deger:1998nm}. Here we will rederive the relevant equation by assuming a simple ansatz for the perturbation. The $L_{-n}$ perturbation is a singlet on the sphere, so we make the ansatz
\be
AdS_3:\quad h_{\mu\nu} ~=~ Mg_{\mu\nu} \,, \qquad\quad S^3:\quad h_{ab} ~=~ Ng_{ab} 
\ee
where for now we take $M$, $N$ (and $P$ to be defined below) to be functions of $t,y,r$ only; we will further restrict this ansatz shortly.
The gauge field on the sphere remains $F^{(3)}_{abc}=2\epsilon_{abc}$ since $\int_{S^3} F^{(3)}$ is a fixed number (the charge on the sphere). Let the field strength on the $AdS$ be perturbed as
\be
F^{(3)}_{\mu\nu\lambda} ~=~ 2(1+P)\epsilon_{\mu\nu\lambda} \,. 
\ee
We look for a perturbation that satisfies $F^{(3)}=*F^{(3)}$, which gives the constraint
\be
(F^{(3)})^2_{S^3} ~=~ -(F^{(3)})^2_{AdS_3} \,.
\label{atwo}
\ee
In the solution (\ref{aone}) we have $(F^{(3)})^2_{S^3}=24$. Under the perturbation above we get $(F^{(3)})^2_{S^3}=24(1-3N)$. On the $AdS_3$, the solution (\ref{aone}) gives ($F^{(3)})^2_{AdS_3}=-24$. Under the perturbation this changes to $(F^{(3)})^2_{AdS_3}=-24(1+2P-3M)$. Thus the condition (\ref{atwo}) gives
\be
-3N ~=~ 2P-3M \,.
\label{athree}
\ee
Now we turn to the Einstein equation $R_{\mu\nu}={1\over 4}F^{(3)}_{\mu\kappa\lambda}(F^{(3)})_\nu{}^{\kappa\lambda}$. For a generic perturbation $h_{AB}$, the change in the Ricci tensor is given by
\be \label{eq:RicciT}
\delta R_{AB} ~=~ -\h[h_{AB;C}{}^C+h^C_{\;C;AB}-h_{AC;B}{}^C-h_{BC;A}{}^C]
\ee
so we have for the perturbation under consideration
\be
\delta R_{\mu\nu} ~=~ -\h[h_{\mu\nu;\lambda}{}^\lambda+h^A_{\;A;\mu\nu}-h_{\mu\lambda;\nu}{}^\lambda-h_{\nu \lambda;\mu}{}^\lambda] \,.
\ee
We have $h^A_{\;A}=3(M+N)$. The stress tensor in the solution (\ref{aone}) is ${1\over 4} F^{(3)}_{\mu\kappa\lambda}(F^{(3)})_\nu{}^{\kappa\lambda}=-2 g_{\mu\nu}$. Under the perturbation this changes to $-2(1+2P)(1-2M) g_{\mu\nu}$. Then the Einstein equation gives 
\be
-\h[\square M g_{\mu\nu}+M_{;\mu\nu}+3N_{;\mu\nu}] ~=~ -2(2P-2M)g_{\mu\nu} \,.
\label{afour}
\ee
Only the terms $M_{;\mu\nu}+3N_{;\mu\nu}$ are not proportional to $g_{\mu\nu}$. We thus set 
\be
M+3N ~=~ 0 \,.
\ee
Then (\ref{athree}) gives 
\be
P ~=~ 2M
\ee
and (\ref{afour}) becomes
\be
(\square-8)M ~=~ 0 \,.
\ee
We seek a solution to this equation of the form $M=f(r)e^{-inv}$ where $v=t-y$. The solution regular at $r=0$ is 
\bea
M&=& A {r^n(n+1+2r^2)\over (1+r^2)^{n\over 2} }e^{-inv} \,
\eea
for constant $A$. Restoring the primes on the coordinates and then going back to the original coordinates using \eq{eq:primed_coords}, this becomes
\bea
M&=& A \left(\frac{r^2}{r^2+a^2} \right)^\frac{n}{2} \left( n+1 + \frac{2r^2}{a^2} \right) e^{-in\frac{v}{R_y}} \,.
\eea

\section{Interior solution for $J^{(3)}_{-n}$ perturbation} \label{app:Sphere}

We now solve for the interior solution which matches onto the perturbation from the sphere diffeomorphisms.
To do so we first write down the equations satisfied by the perturbation. We then rewrite these equations in terms of differential forms. Finally, we reduce the resulting equations to a single second order equation, which can be solved in closed form.

\subsection{The equations satisfied by the perturbation} \label{sec:J_n_int}

Following \cite{Deger:1998nm,Mathur:2001pz}, we make the ansatz
\be
h_{a\mu}(x, y) ~=~ K_\mu(x) Y_a(y), \quad C_{\mu a}(x, y) ~=~ Z_\mu(x) Y_a(y) \,.
\label{ansatza}
\ee
The vector spherical harmonic $Y$ has has contravariant components
\be
Y^\theta ~=~ 0, \quad Y^\psi ~=~ 1, \quad Y^\phi ~=~ -1
\ee
and satisfies
\be
Y_{a;b}{}^b ~=~ -2Y_a \,.
\ee
The relevant component of the Einstein equation is $R_{a\mu}={1\over 4}F^{(3)}_{aAB}F^{(3)}{}_{\!\mu}{}^{AB}$. With the ansatz (\ref{ansatza}), 
we obtain
\be
R_{a\mu} ~=~ -\h [K_{\mu; \nu}{}^\nu -K_{\nu;\mu}{}^\nu-4K_\mu]Y_a \,,
\ee
\be
{1\over 4}F^{(3)}_{aAB}(F^{(3)})_\mu^{~AB} ~=~ \h[\epsilon_{a}{}^{bc}F^{(3)}_{\mu bc}+F^{(3)}_{a\nu\lambda}\epsilon_{\mu}{}^{\nu\lambda}] ~=~ \h[-4 Z_\mu - (Z_{\nu; \lambda}-Z_{\lambda; \nu})\epsilon_{\mu}{}^{\nu\lambda}]Y_a
\ee
where we have used
\be
F^{(3)}_{\mu\nu\lambda}~ ~=~ ~2\epsilon_{\mu\nu\lambda}, \qquad F^{(3)}_{abc} ~ ~=~ ~2\epsilon_{abc} \,.
\ee
Thus the Einstein equation becomes
\be
[K_{\mu; \nu}{}^\nu -K_{\nu;\mu}{}^\nu-4K_\mu]+[-4 Z_\mu - (Z_{\nu; \lambda}-Z_{\lambda; \nu})\epsilon_{\mu}{}^{\nu\lambda}]~=~0 \,.
\label{eqone}
\ee
The relevant component of the self-duality condition is $(*F^{(3)})_{\mu\nu a}=F^{(3)}_{\mu\nu a}$, which gives
\be
\epsilon_{\mu\nu\lambda}[2K^{\lambda}+2Z^{\lambda}]~=~(Z_{\nu; \mu}-Z_{\mu; \nu}) \,.
\label{eqtwo}
\ee

\subsection{Rewriting the equations in 3D form language}

We now denote by $K, \, Z$ the 1-forms $K_\mu, \, Z_\mu$ and we define the field strengths 
\be
F ~=~ dK \,, \qquad  G~=~dZ \,. 
\ee
In this section $*$ denotes the 3D Hodge dual. The equations (\ref{eqone}), (\ref{eqtwo}) become
\be
*d*F-4K-4Z+2*G~=~0 \,,
\label{aaone}
\ee
\be
2(*K+*Z)~=~G \,.
\label{aatwo}
\ee
Noting that $**K=-K$ and similarly for $Z$, we can write (\ref{aatwo}) as $2(K+Z)=-*G$. Substituting this in (\ref{aaone}) we get
\be
*d*F+4*G~=~0 \quad \Rightarrow \quad d*F+4 \,dZ~=~0 \,.
\ee
This has the solution $Z=-{1\over 4}*F$ up to gauge transformations $Z\r Z+d\Lambda_Z$. We write this as
\be
F~=~4*Z \,.
\label{aathree}
\ee
Then from (\ref{aatwo}) we get 
\be
K~=~-Z-\h *G \,.
\label{aafour}
\ee
Applying the operator $d$ to both sides of this relation gives $ F=-G-\h d*G$. Using (\ref{aathree}) we get a relation involving $Z$ alone: $4*Z=-G-\h d*G$. In terms of $Z$, this is $*d*dZ+2*dZ-8Z=0$, which can be written as
\be
(*d-2)(*d+4)Z ~=~ 0 \,.
\ee
This splits into a pair of equations
\be
(*d-2)Z~=~0 \,, \quad {\rm or}\quad (*d+4)Z~=~0 \,.
\label{aaseven}
\ee
Below we will solve the general equation
\be
*dZ+\alpha Z~=~0 \,.
\label{aafive}
\ee
We will see that the solution we need corresponds to the choice $\alpha=-2$. For this value of $\alpha$ we note from (\ref{aafour}) that
\be
K~=~-2Z \,.
\label{qesix}
\ee

\subsection{Solving the equations}\label{aeq}

We wish to solve the equation (\ref{aafive}) with an ansatz of the form
\be
Z_v ~=~ f(r) e^{-in v}, \qquad 
Z_u ~=~ q(r) e^{-in v}, \qquad 
Z_r ~=~ h(r) e^{-in v} \,.
\ee
Then the field strength $G=dZ$ has nonzero components
\be
G_{rv} ~=~ (f'+in h)e^{-in v}, \qquad G_{ru} ~=~ q' e^{-in v}, \qquad 
G_{uv} ~=~ in qe^{-in v} \,.
\ee
The $v,\,u$ and $r$ components of the equation $*G=-\alpha G$ then give the equations
\bea
{1\over 2r}\left[-q'+({2r^2+1})(f'+in h)\right] &=& -\alpha f \,, \label{aa1}\\
-{1\over 2r}\left[({2r^2+1})q'-(f'+inh)\right] &=& -\alpha q \,, \label{aa2}\\
{2in  \over r(r^2+1)}q &=& -\alpha h \,. \label{aa3}
\eea
The last relation (\ref{aa3}) can be immediately solved to give
\be
h ~=~ -{2in  \over \alpha r(r^2+1)}q \,.
\ee
Rearranging (\ref{aa1}) we get
\be
f'+inh  ~=~  {-2r\alpha f+q'\over 2r^2+1}
\ee
which we substitute in (\ref{aa2}) to get 
\be
f ~=~ -{2r(r^2+1)\over \alpha}q'+(2r^2+1)q \,.
\ee
Substituting $f$ back in (\ref{aa1}) we get
\be
r^2(r^2+1)^2q''+r(r^2+1)(3r^2+1)q'-\left[\alpha(\alpha+2)r^2(r^2+1)+n^2\right]q ~=~ 0 \,.
\label{aaeight} 
\ee

\subsection{Selecting a solution}\label{secqw}

We now want to solve (\ref{aaeight}). The equation for the perturbation (\ref{aaseven}) allowed the values $\alpha=-2, \alpha =4$. The solution which matches on to the perturbation we construct in the neck is the $\alpha=-2$ solution, which has the form
\be
q ~=~ C \left( {r^2\over 1+r^2} \right)^{n\over 2} + D \left( {1+r^2\over r^2} \right)^{n\over 2} \,.
\ee
We see that regularity at the origin requires $n$ to be an integer. For $n>0$, we set $D=0$, getting $q=C ({r^2\over 1+r^2})^{n\over 2}$. 

So the solution is 
\bea
Z_v &=&  C\Big ( {r^2\over r^2+1}\Big ) ^{n\over 2} \Big ( n+1+2r^2\Big ) e^{-i n v}   \,,  \cr
Z_u &=&  C\Big ( {r^2\over r^2+1}\Big ) ^{n\over 2} e^{-i n v}    \,,  \\
Z_r &=&  C{in\over r(r^2+1)} \Big ( {r^2\over r^2+1}\Big ) ^{n\over 2} e^{-i n v}    \,. \nnm
\eea
Restoring the primes on the coordinates and then going back to the original coordinates using \eq{eq:primed_coords}, we get
\bea
Z_v &=&  C \, a \Big ( {r^2\over r^2+a^2}\Big ) ^{n\over 2} \left( n+1 + \frac{2r^2}{a^2} \right) e^{-i n {v\over R_y}}   \,,  \cr
Z_u &=&  C \, a \Big ( {r^2\over r^2+a^2}\Big ) ^{n\over 2} e^{-i n {v\over R_y}}    \,, \label{qeten_app} \\
Z_r &=&  C \, in {a^2 Q\over r(r^2+a^2)} \Big ( {r^2\over r^2+a^2}\Big ) ^{n\over 2} e^{-i n {v\over R_y}}  \,, \nnm
\eea
and from (\ref{qesix}) we recall that for $\alpha=-2$ we have
\bea
K_\mu &=& -2Z_\mu \,.
\eea

\section{Interior solution for $J^i_{-n}$ perturbation} \label{appcq}

We now solve for the interior solution which matches onto the perturbation from the torus diffeomorphism.
We make the ansatz
\be
h_{\mu z} ~=~ K_\mu, \quad C_{\mu z} ~=~ Z_\mu \,.
\ee
We define the field strengths
\be
F_{\mu\nu} ~=~ \p_\mu K_\nu-\p_\nu K_\mu, \quad G_{\mu\nu} ~=~ \p_\mu Z_\nu-\p_\nu Z_\mu
\ee
The relevant component of the Einstein equation is $R_{z\mu}={1\over 4}F^{(3)}_{zAB}F^{(3)}_\mu{}^{AB}$ which gives
\be
F_{\mu\nu}{}^{;\nu} ~=~ {1\over 2}G_{\mu_1\mu_2}(\bar F^{(3)})^{\mu_1\mu_2}{}_\mu
\ee
where $\bar F^{(3)}$ is the background value of $F^{(3)}$.
The field equation $F^{(3)}_{z\mu A}{}^{;A}=0$ gives
\be
G_{\mu\nu}{}^{;\nu}-\h F_{\mu_1\mu_2}\bar F^{(3)\mu_1\mu_2}{}_ \mu ~=~ 0 \,.
\ee
Again using $*$ for the 3D Hodge dual, the Einstein equation in terms of forms is
\be
*d*F ~=~ -2 *G
\ee
and the field equation is
\be
*d* G ~=~ -2 *F \,.
\ee
Adding and subtracting these equations, we get two decoupled equations
\bea
*d*(F+G) ~=~ -2 *(F+G)           ~~& \Rightarrow &~~ d*(F+G) ~=~ -2(F+G) \cr
{}*d*(F-G) ~=~ \phantom{-} 2*(F-G) ~~& \Rightarrow &~~ d*(F-G) ~=~ \phantom{-} 2(F-G)
\eea
These equations are of the form (\ref{aafive}), i.e. of the form $d*H+\alpha H=0$ 
for a 2-form $H$. If $H=dA$, then we get $d*dA+\alpha dA=0$. This gives
\be
*dA+\alpha A ~=~ 0
\ee
up to gauge transformations $A\r A=d\Lambda_A$. Thus the 1-forms $K+Z, K-Z$ satisfy (\ref{aafive}) with $\alpha = 2, -2$ respectively. 

We find that the solution which will matches to the solution in the outer region is obtained by choosing $K+Z=0$. Then the equation for $K-Z=-2Z$ is the same as the equation for $Z$ solved in section \ref{secqw}, namely $(*d-2)Z=0$, so the solution for $Z$ has the same form as for the sphere case:
\bea
Z_v &=&  D \, a \Big ( {r^2\over r^2+a^2}\Big ) ^{n\over 2} \left( n+1 + \frac{2r^2}{a^2} \right) e^{-i n {v\over R_y}}   \,,  \cr
Z_u &=&  D \, a \Big ( {r^2\over r^2+a^2}\Big ) ^{n\over 2} e^{-i n {v\over R_y}}    \,, \label{qeten_app_z} \\
Z_r &=&  D \, in {a^2 Q\over r(r^2+a^2)} \Big ( {r^2\over r^2+a^2}\Big ) ^{n\over 2} e^{-i n {v\over R_y}}  \, \nnm
\eea
for some constant $D$. This solution has $K+Z=0\,,$ so $K$ is given by 
\bea
K_\mu &=& -Z_\mu \,.
\eea

\end{appendix}

%\newpage

\providecommand{\href}[2]{#2}\begingroup\raggedright\endgroup

\end{document}